\documentclass[aps,twocolumn,floats,nofootinbib,11pt,tightenlines]{revtex4-2}

\setcitestyle{super}
\usepackage{newtxtext,newtxmath}
\usepackage[centering,hmargin=2cm,vmargin=2.5cm]{geometry}

\usepackage{mathtools}
\usepackage{siunitx}
\usepackage{amsmath}
\usepackage{graphicx}
\usepackage[T1]{fontenc}
\usepackage[utf8]{inputenc}
\graphicspath{{images/}}
\usepackage{color}
\usepackage[pdfstartview=FitH,
            breaklinks=true,
            bookmarksopen=false,
            bookmarksnumbered=true,
            colorlinks=true,
            linkcolor=black,
            citecolor=black,
            urlcolor=black,
            pdftitle={Self and nonself},
            pdfauthor={Jonathan Levine},
            pdfsubject={}
            ]{hyperref}
\newcommand{\B}{\boldsymbol}
\newcommand{\ud}{\mathrm{d}}

\def\(({\left(}
\def\)){\right)}                       
\def\[[{\left[}
\def\]]{\right]}

\begin{document}

\title{How different are self and nonself?}

\author{Andreas Mayer,$^{a,b,\dag}$ Jonathan A. Levine,$^{c,d,h,\dag}$ Christopher J.~Russo,$^{a,e}$ Quentin Marcou,$^{f}$ William Bialek,$^{a,g}$ Benjamin D.~Greenbaum$^{h,i}$}

\affiliation{$^a$Joseph Henry Laboratories of Physics and Lewis–Sigler Institute for Integrative Genomics, Princeton University, Princeton NJ 08544 USA\\
$^b$Division of Infection and Immunity and Institute for the Physics of Living Systems, University College London, London NW3 2PP, UK \\
$^c$Tri-institutional PhD Program in Computational Biology and Medicine, Weill Cornell Medicine, New York, NY USA\\
$^d$Laboratory of Lymphocyte Dynamics, The Rockefeller University, New York, NY USA\\
$^e$Program in Biophysical Sciences, The University of Chicago, Chicago IL 60637 USA \\
$^f$Faculté de Médecine La Timone, Aix-Marseille Université, 13005 Marseille, France \\
$^g$Center for Studies in Physics and Biology, Rockefeller University, New York, NY 10065 USA
$^h$Computational Oncology, Department of Epidemiology and Biostatistics,  Memorial Sloan Kettering Cancer Center, New York, NY 10065 USA \\
$^i$Physiology, Biophysics \& Systems Biology, Weill Cornell Medicine, Weill Cornell Medical College, New York, NY 10065 USA\\
$\dag$ These authors contributed equally}

\date{\today}

\begin{abstract}

Biological and artificial networks routinely make reliable distinctions between similar inputs, and the rules for making these distinctions are learned. In some ways, self/nonself discrimination in the immune system is similar, being both reliable and (partly) learned through thymic selection. In contrast to other examples, we show that the distributions of self and nonself peptides are nearly identical but strongly inhomogeneous. Reliable discrimination is possible only because self-peptides are a particular finite sample drawn out of this distribution, and T cells can target the “spaces” in between these samples. In conventional learning problems, this would constitute overfitting and lead to disaster. Here, the strong inhomogeneities imply instead that the immune system gains by targeting peptides which are similar to self, with maximum sensitivity for sequences just one or two substitutions away. This prediction from the structure of the underlying distribution in sequence space agrees, for example, with the observed responses to mutation derived cancer neoantigens.

\end{abstract}

\maketitle

A basic task of the immune system is to distinguish self from nonself. More specifically, cytotoxic T cells should respond to relatively short peptides from foreign antigens and not respond to innocuous peptides, such as those synthesized by the organism itself. One might expect this task is easier when the differences between peptides are larger, and the targeted peptides are therefore less similar to self. Yet, recent work in cancer immunology has shown that cancer neoantigens that differ from self by just a single amino acid substitution can still generate strong immune responses.\cite{Schumacher2015} More generally, T cells are often capable of recognizing peptides that are close to self. The strength of such immune responses can exert a selective pressure on tumors, to which they adapt by activating immune checkpoints, thereby attenuating immune cell function. Inhibiting such checkpoint blockades has been the basis of transformative immunotherapies. \cite{Schumacher2015, Matsushita2012, Luksza2017, Ribas2018, Wells2020, Luksza2022, rojas2023personalized} \\
\indent T cells learn to distinguish between self and nonself, at least to some extent, through the process of thymic selection.  Like for other learning problems\cite{Mehta2019,Carleo2019} this process requires some degree of generalization,\cite{Butler2013,Legoux2015,Davis2015,yu2015clonal,Wortel2020} because it is not possible for each candidate T cell to be tested against all possible self-peptides.  In a usual learning problem, examples are drawn from a distribution, and it is essential to capture features of this underlying distribution and not “overfit” to the particular random samples that one has seen.  Here we will show that for the self/nonself distinction in peptides the opposite is the case, and overfitting is both allowed and necessary. Our analysis allows us to understand why peptides that are close to self are natural targets for the immune system. \\
\indent To approach these issues we need a model for the relevant distributions.  We use a maximum entropy approach to analyze the occurrence of amino acid motifs that characterize peptides across all domains of life. Maximum entropy models have origins in statistical physics\cite{jaynes1957information} but have also been used in machine learning, particularly in natural language processing.\cite{berger1996maximum} These “small language models” are ideal for capturing the distribution of the short peptides presented by the class I major histocompatibility complex (MHC-I) in an interpretable fashion. As generative models they easily generalize to the large language models (LLMs) which currently underly unsupervised generative transformer models\cite{radford2019language} and are beginning to be applied to more complex computational proteomic tasks.\cite{madani2023large,nagano2025contrastive} \\
\indent Using maximum entropy models, we mathematically characterize the statistical structure of the self/nonself discrimination problem facing cytotoxic T cells. This structure is defined by the distribution of sequences found in the ~9-mer peptides presented to the immune system by MHC-I. The key idea is to build models that match specific features of the data exactly, but otherwise have as little structure as possible.  This strategy has produced very accurate models for e.g., the joint patterns of activity in networks of neurons,\cite{Schneidman2006, Maoz2020,Meshulam2021, meshulam2024statistical} correlated variations of amino acids in protein families\cite{Lapedes1998, Bialek2007,Weigt2009, Cocco2018, Marks2011}, and fluctuating flight velocities in flocks of birds.\cite{Bialek+al2012}  Importantly, the maximum entropy method allows an incremental incorporation of constraints, allowing for a more precise description as more features of the data are added.\\
\indent Here we show that, across eukaryotes, such models are unable to distinguish between hosts and pathogens. In other words, host and pathogen proteomes are described by the same underlying language model, reflecting their amino acid statistics. This is true even when specific MHC haplotypes are considered, focusing only on filtered subsets of presented peptides. We characterize this landscape and find that encountering a peptide that is only one mutation away from self, which is often the case for cancer neoantigens, is not an unusual problem for the immune system to solve when fighting pathogens. On the contrary, the immune recognition of such “close to self” peptides might be evolutionarily favorable.  
\section*{Results}
\subsection*{Proteome Biases Reduce Diversity of Peptides Beyond MHC Presentation}
 \begin{figure*}[t]
     \includegraphics[width=\textwidth]{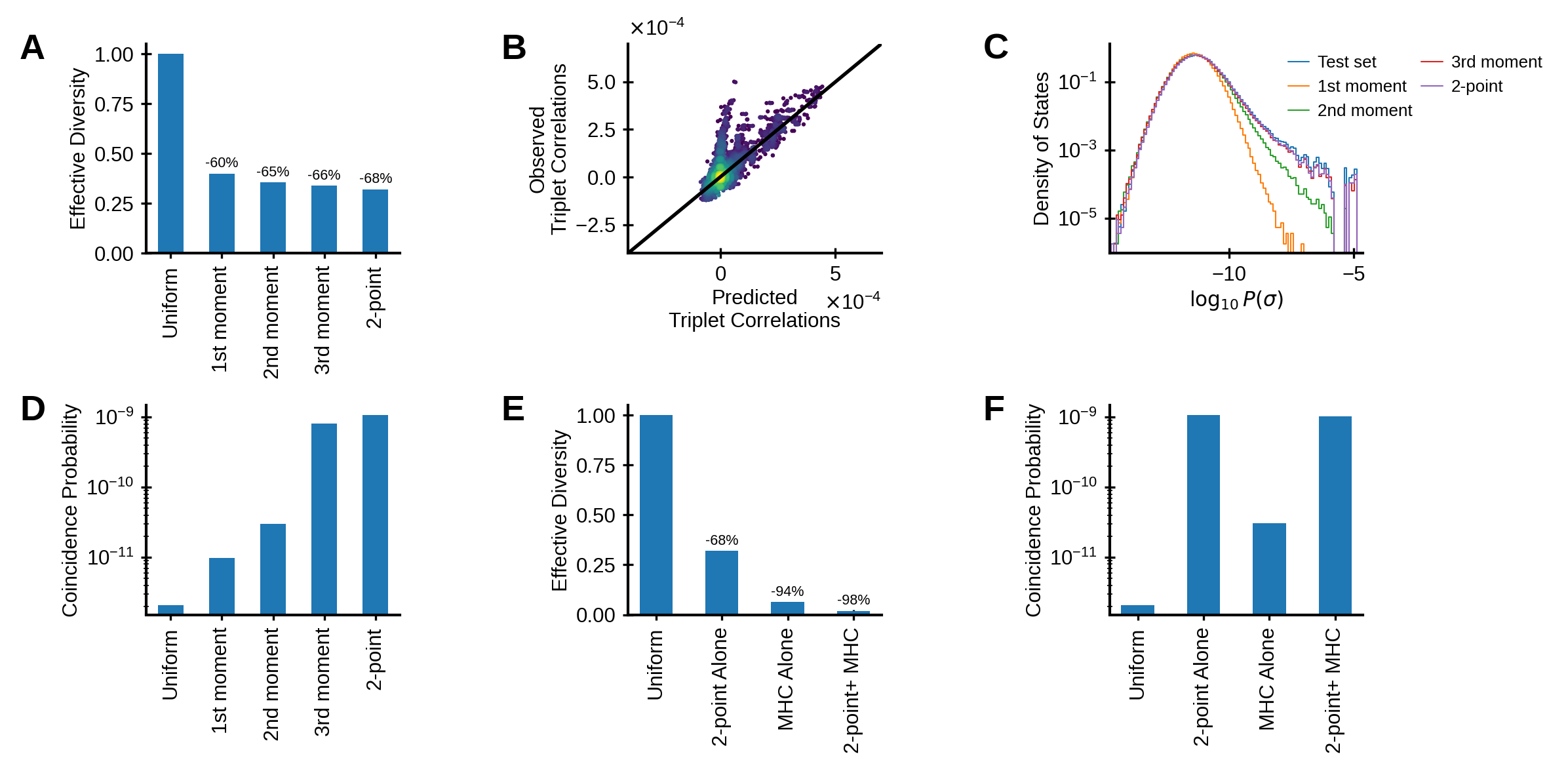}
        \caption{{\bf Maximum entropy models predict peptide statistics.} \\  (A) Entropy decreases as we add more constraints on average statistics across sites (moment terms) and correlations between specific sites (2-point term), corresponding to successive terms in the energy (Eq. 3). (B) Predicted vs observed triplet correlations among amino acids at different sites. (C) Probability density of (log) probability in the full model. We compare ensembles of sequences drawn from the series of models in (A), and from the test set data. (D) Probability that two randomly chosen sequences are identical, computed across ensembles drawn from the series of models in (A). (E) Entropy reduction from the full model alone, MHC-I presentation alone, and a joint distribution of the model and MHC-I presentation together. (F) Probability that two randomly chosen sequences are identical, computed across ensembles drawn from the series of models in (E). Effective diversity is calculated as $D=exp(S)/20^9$, where $S$ is the Shannon entropy. $D$ can be interpreted as the Shannon diversity (also known as perplexity) relative to that of a uniform distribution of 9-mers. Effective diversity and coincidence for “MHC alone” and “2-point+MHC” (E-F) are calculated as the average across independent models for the top 500 most common haplotypes (see Figure S2). }
 \end{figure*}
 
The twenty amino acids are used with unequal probabilities\cite{Lehmann2016} in vertebrate proteomes and the proteomes of large groups of infectious viruses, bacteria, or parasites\cite{UniprotConsortium2021}(\textbf{Figure S1}, for details of the databases we use see \textbf{Appendix A}). Correlations between neighboring amino acids are weak, but they extend across a very long distance, far beyond the length of the relevant peptides (\textbf{Figure S1B}).  For k-mers with k < 4, the size of the human proteome allows for sampling of all possible sequences, and we can estimate the probability distribution with reasonable control over errors. However, there are too many possible 9-mers, and a statistical model is required to define the underlying distribution. \\
\indent Mathematically, we choose a probability distribution of 9-mer sequences, $\sigma$, that maximizes the Shannon entropy:
\begin{equation}
	 S[P(\sigma)] = - \sum_{\sigma}P(\sigma)\log P(\sigma)
\end{equation}
subject to the normalization constraint $\sum_{\sigma} P(\sigma)=1$, and constraints that enforce the equality of modelled and empirical observables. \\
\indent We begin by matching global statistical features of observed sequences: first the mean number of each type of amino acid that appears in the peptides (\textbf{first moment}), then the covariances in these numbers (\textbf{second moment}), and then the asymmetry of their probability distribution (\textbf{third moment}).  Finally, we include the fact that correlations between pairs of amino acids depend on the distance between them within the 9-mer (\textbf{2-point}).  Each of these additional constraints is improving our model because each contributes to lowering the entropy (\textbf{Figure 1A}).  

Formally, we can define $s_{\rm i}^\alpha =1$ if the amino acid at site $\rm i$ is of type $\alpha$,  and $s_{\rm i}^\alpha =0$ otherwise; ${\rm i} = 1,\, 2,\, \cdots ,\, 9$ runs along the length of the peptide and $\alpha = 1,\, 2,\, \cdots ,\, 20$ over the amino acids.  We will refer to the entire sequence as ${\boldsymbol  \sigma} = \{s_{\rm i}^\alpha\}$. 
\begin{widetext}
 It is useful to count the number of amino acids of each type that appear in the peptide, $n^\alpha = \sum_{\rm i} s_{\rm i}^\alpha$. With these definitions, the maximum entropy model described above takes the form:
\begin{equation}
P({\boldsymbol  \sigma}) = {1\over Z} \exp\left[ - E({\boldsymbol \sigma})\right] , \label{model1}
\end{equation}
where the "energy", $E$, depends on the variables $s_{\rm i}^\alpha$ and $n^\alpha$ and their moments:

\begin{equation}
E({\boldsymbol \sigma}) = \sum_{\alpha=1}^{20} \lambda_1^\alpha n^\alpha + \sum_{\alpha,\beta=1}^{20} \lambda_2^{\alpha\beta} n^\alpha n^\beta + \sum_{\alpha,\beta,\gamma=1}^{20} \lambda_3^{\alpha\beta\gamma} n^\alpha n^\beta n^\gamma + {1\over 2}\sum_{{\rm i},{\rm j} = 1}^9 \sum_{\alpha ,\beta=1}^{20} J^{\alpha\beta}_{\Delta}  s_{\rm i}^\alpha s_{\rm j}^\beta 
\end{equation}
\end{widetext}
All coefficients must be adjusted so that the predictions of the model match observed features of the data\cite{Ackley1985}: $\lambda_1^\alpha$ is fixed by matching the mean number of amino acids of type $\alpha$, $\lambda_2^{\alpha\beta}$ by matching the covariance in numbers of amino acids of types $\alpha$ and $\beta$, $\lambda_3^{\alpha\beta\gamma}$ by matching the third moments of these numbers, and the matrix $J^{\alpha\beta}_{\Delta}$, where $\Delta=|{\rm i}-{\rm j}|$, must be adjusted to match the distance dependence of the pairwise correlations. The constant (partition function) $Z$ serves to normalize the distribution, so that $Z = \sum_{\boldsymbol\sigma}  \exp\left[ - E({\boldsymbol \sigma})\right]$ and can be evaluated numerically by thermodynamic integration\cite{Marchi2019b} (\textbf{Appendix B}). We emphasize that once we match these coefficients to measured features of the proteome, there are no free parameters that can be adjusted. \\
\indent This family of models makes accurate predictions for higher order statistical properties of the distribution of human 9-mers. As an example, we can compute the correlations among triplets of amino acids at particular sites:
\begin{equation}
C_{\rm ijk}^{\alpha\beta\gamma} = \langle \delta s_{\rm i}^\alpha \delta s_{\rm j}^\beta \delta s_{\rm k}^\gamma \rangle ,
\label{triplet}
\end{equation}
with $\delta s_{\rm i}^\alpha =  s_{\rm i}^\alpha - \langle  s_{\rm i}^\alpha\rangle$.
While small, the predicted correlations largely agree with those in the observed data (\textbf{Figure 1B}).  More subtly, the model assigns a probability of occurrence to every possible 9-mer, and we can calculate the distribution of this (log) probability (\textbf{Figure 1C}).  As we constrain more features, this distribution grows a “heavy tail” of relatively high probability sequences that are also in real data. To calibrate this effect, we note that there are nearly $10^{12}$ possible 9-mers, so the prediction that any sequence occurs with probability above $10^{-7}$ reflects a startling 10,000-fold enrichment (or more) over random sequences, and many of these predicted sequences are indeed found in real data. A corollary of this prediction is that two 9-mer sequences chosen at random are thousands of times more likely to be the same than if their individual amino acids were chosen at random (\textbf{Figure 1D}). Large effects on the coincidence probability coexist with relatively small effects on the entropy precisely because the distribution is strongly non-uniform, which has important implications for the immune response. \\

\indent To understand the potential biological importance of the evolutionary constraints uncovered by our modelling approach, we compared their impact to the well-known constraints imposed by MHC-I presentation, which restricts peptides to those with specific amino acids at anchor sites. Using a previously derived distribution of human HLA haplotypes which accounts for linkage,\cite{hoyos2022fundamental} we created a new model for each haplotype describing the filtered distribution of peptides in the proteome that are predicted to be presented by any allele in that haplotype. Using these filtered distributions for each of the top 500 most common haplotypes (\textbf{Figure S2, Appendix C}) we compared how the maximum entropy model and MHC-I restriction each contribute to reducing the effective diversity of peptides presented to the immune system (\textbf{Figure 1E}).  On average, MHC-I presentation alone reduced the effective diversity by $94\%$ relative to the uniform distribution, compared to $68\%$ with the full 2-point maximum entropy model. Combining the constraints captured by the maximum entropy model with the restriction of MHC-I presentation acted additively to reduce diversity by a total of $98\%$ relative to uniform, implying both models restrict non-redundant information. \\
\indent Furthermore, we found that the existence of the “heavy tail” of highly probable peptides in our model increased the probability of coincidence by $10^3$ compared to an increase by $10^1$ with MHC-I presentation alone (\textbf{Figure 1F}). Additionally, the percentage of peptides that were classified as binders did not differ significantly between the family of models and did not vary significantly between haplotypes (\textbf{Figure S2B}). Thus, while MHC-I presentation determines which peptides can be presented, our maximum entropy model captures independent information on the most common sequence motifs, which are likely to be more biologically relevant for immune recognition. Put another way, MHC presentation largely contributes to the reduction in the Shannon entropy of the resulting distribution (\textbf{Figure 1E}), while our model highlights the reduction in Renyi entropy, as reflected in the coincidence statistics (\textbf{Figure 1F}). This suggests that the biophysical constraints on peptide distributions might be an important additional consideration in understanding the evolution of immune recognition. Indeed, we found that these constraints also have an impact on the observed overlap between self and nonself proteomes.

\begin{figure*}[t]
    \begin{center}
        \includegraphics[width=\textwidth]{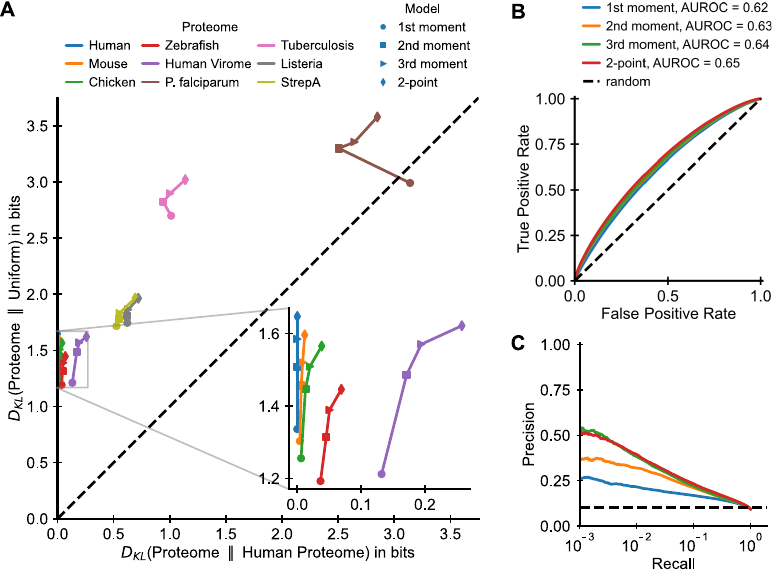}
    \end{center}
    \caption{
        {\bf Divergence between peptides from different proteomes.} (A) Kullback-Leibler ($KL$) divergences between peptide distributions of different pathogen proteomes relative to human host peptides and relative to a uniform distribution over all peptides. For each proteome we show the statistical distance calculated according to a nested set of models including a different number of constraints. The inset shows a zoom on the set of proteomes close to the human statistics. (B, C) Performance of the models as classifiers, where we assign peptides as self or nonself (viral) based on the likelihood ratio $L = \log[P_{\rm nonself} (\sigma)/ P_{\rm self} (\sigma)]$. (B) If we choose a nonself viral peptide at random, there is some probability that it will be identified correctly (true positive), and if we choose a self-peptide and random there is some probability it will be identified incorrectly as nonself (false positive). We can trade these probabilities against one another by changing the threshold L at which we make the decision. (C) If self-peptides are in 10-fold excess, there is a probability that some random peptide will be classified correctly (precision), and this will “catch” a certain fraction of the nonself peptides (recall). Note that if peptides are assigned self/nonself at random, we will be right $10\%$ of the time since self is in 10-fold excess (dashed line).
    }
\end{figure*}

\subsection*{Comparisons between self and nonself proteomes}

Having found a class of models that describes the distribution of human 9-mer sequences quantitatively, we can ask whether this distribution differs between humans and other organisms, or between humans and pathogens. We build maximum entropy models for 9-mers coming from a variety of proteomes and compare the resulting distributions. The natural measure of difference between distributions is the Kullback-Leibler divergence ($D_{KL}$), which measures the average evidence (or log-likelihood ratio) that a single peptide belongs to one distribution rather than the other.\cite{cover1999elements} For each proteome we show the KL divergences relative to both the human peptide distribution and relative to a uniform distribution over all peptides (\textbf{Figure 2A}).\\
\indent The distributions of 9-mers in these proteomes are very different from the uniform distribution, but quite similar to one another, and to the statistics of the human proteome (\textbf{Figure 2A, Figure S3}). As we add successive constraints, we find that not just amino acid frequencies, but also covariances and higher-order statistics are largely shared across proteomes. For instance, the proteomes of bacteria and their eukaryote hosts all differ more from the uniform distribution ($D_{KL} > 1.0 \, \rm bit$), than from each other ($D_{KL} \lessapprox 1.0 \, \rm bit$), and the maximal $D_{KL}\approx 1.0 \, \rm bit$ still makes individual 9-mers only weakly predictive. The largest divergence is found for the parasite \textit{Plasmodium falciparum}, an organism known to have an unusually AT-rich genome.\cite{Hamilton2017} However, the \textit{P. falciparum} proteome is still closer in statistical distance to the human proteome than it is to a uniform model. Most strikingly, if we compare a collection of human viral proteomes with the human proteome, we see $D_{KL} \approx 0.2 \, \rm bits$, indicating that individual viral peptides are on average almost indistinguishable from self-peptides. A consequence of the small $D_{KL}$ is that an immune system trained on general features of these distributions can recognize individual peptides from viruses only if it is willing to accept a nearly comparable false positive rate, corresponding to autoimmune recognition (\textbf{Figure 2B}).\\
\indent The trade-off captured by the receiver operating characteristic is independent of the overall probability with which self or nonself peptides are encountered. If, for example, there is a 10-fold excess of self-peptides, we can calculate the probability that any one peptide will be identified correctly (precision), but we can trade this against the fraction of nonself peptides that will be recognized (recall).  As shown by the precision-recall curve (\textbf{Figure 2C}), even if such an immune system catches only one in a thousand nonself challenges (those which “stick out” as being most unlikely to occur in the self-distribution), it still reaches only $\sim 50 \%$ chance of being correct about a random peptide. Moreover, while skewing MHC presentation to more easily discriminate peptides might be advantageous in principle, empirical analysis of the MHC filtered viral peptides shows that while their distributions are also skewed by presentation, the resulting filtered distributions remain highly similar between viral and self-peptides (\textbf{Figure S4A}). In other words, both host and viral peptides are skewed in the same manner: the degree of similarity between the two distributions remains when considering only MHC presented peptides. Consequently, a similar classification analysis between self and viral peptides limited to those that are predicted to be presented on MHC-I shows no performance gain (\textbf{Figure S4B-C}). This suggests that a functional immune system does not distinguish between self and nonself by learning general features of the underlying distributions.

\begin{figure*}[t]
	\begin{center}
		\includegraphics[width=\textwidth]{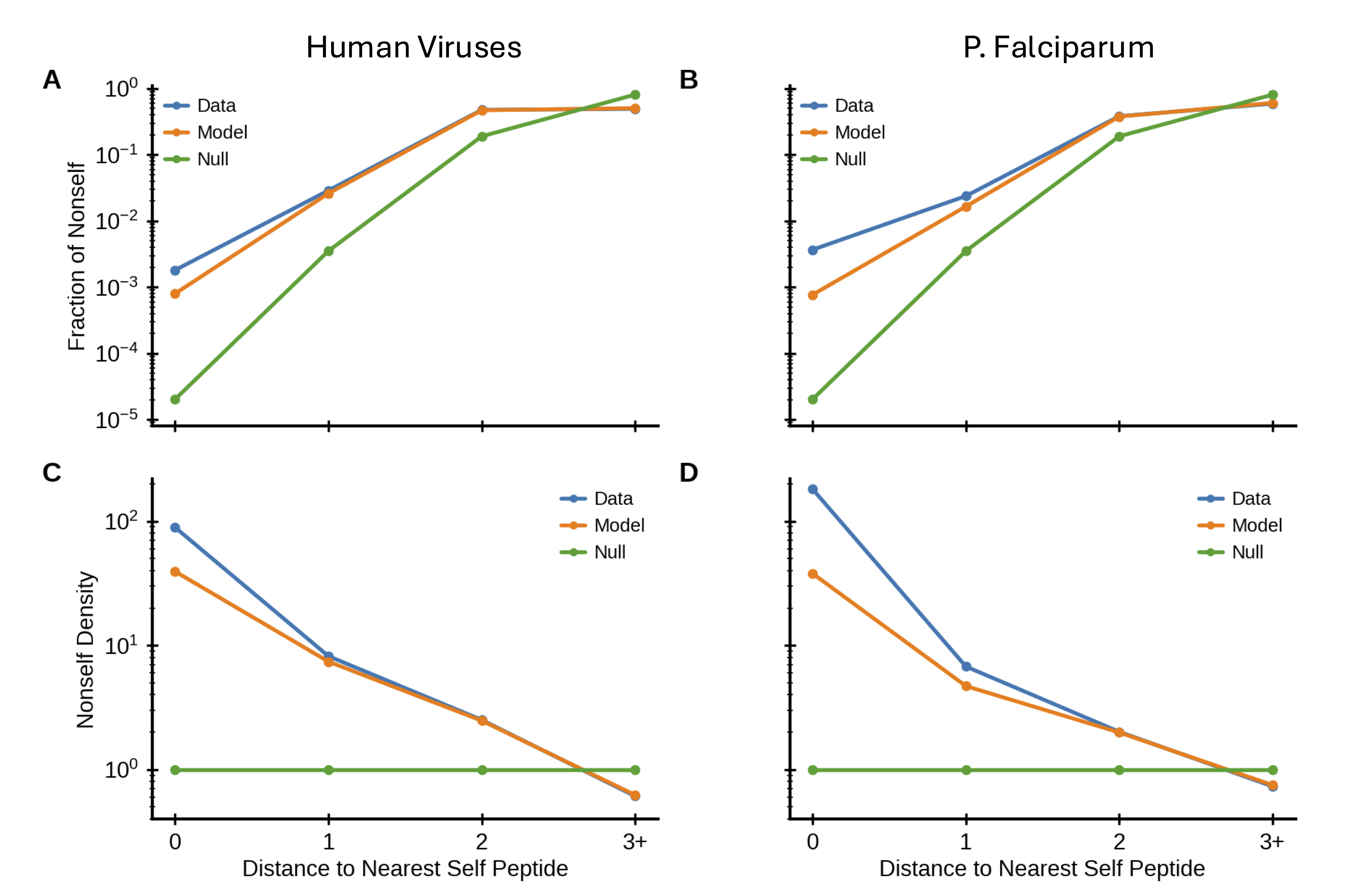}
	\end{center}
    \caption{
        {\bf Many nonself peptides are close to self.} (A-B) Distribution of distances to the nearest self-peptide for peptides from an ensemble of human viruses (A) and Plasmodium Falciparum (B) as found in the data (blue), in predictions (orange) from the full model, and in a null model with a uniform distribution over all 209 9–mers (green). (C-D) Relative density of distances to the nearest self-peptide for peptides from human viruses (C) or P. Falciparum (D). Colors as in (A-B).
        Note: Owing to the size of these proteomes, statistical error bars on these estimates are too small relative to the data to appear on the plots (See Appendix D).  
        }
\end{figure*}

\subsection*{The density of nonself peptides skews close to self}

\indent Due to the discrepancy between the $\sim 10^{12}$ possible 9-mers versus the  $\sim 10^{7}$ realized 9-mers in the human proteome, the immune system can recognize pathogen sequences that fall into the spaces between the learned self-sequences even if they come from the same overall language model. But these gaps are tight: if sequences were random, typical self-peptides would be separated by only three amino acid substitutions. In fact, the distribution is very inhomogeneous, as discussed above, so most of the gaps are even smaller.\\
\indent We can make this geometric picture explicit by asking, for each 9-mer in a nonself proteome, what is the distance to the nearest human sequence.  Looking at an ensemble of human viruses, we find that more than $\sim 0.1 \%$ of these pathogenic sequences are identical with a human peptide, and more than $1\%$ are just one amino acid away (\textbf{Figure 3A}).  These coincidences and near coincidences are 10 – 100 times more frequent than expected if sequences were generated either by a uniformly random null model (\textbf{Figure 3A}) or by a model with shuffled amino acid frequency usages (\textbf{Figure S5}). This shifted nearest-distance distribution is recapitulated with increasing precision as more constraints are added to the model (\textbf{Figure S5}) and near coincidences are captured perfectly by the full model in Eq. (2) (\textbf{Figure 3A, Figure S6}).  The scale of these effects is biologically meaningful: the empirical coincidence probability of $\sim 10^{-3}$ implies that for a viral proteome with $\sim 10^4$ amino acids, exact coincidences are common, whereas they would be rare in the absence of the biases captured in our model. Lastly, we repeated the analyses with peptides from a variety of pathogen proteomes and the same pattern emerged (\textbf{Figure S7A}). Even for the most statistically distinguishable proteome, from the malaria parasite \textit{Plasmodium falciparum}, we again found a large enrichment of near coincidences (\textbf{Figure 3B}), showing that these findings generalize even to comparatively more dissimilar proteomes.\\
\indent There are more pathogenic sequences within one amino acid of a human sequence than perfect coincidences in part because there are 9 x 19 = 171 ways to be at distance one, but only one way to be at distance zero.  Normalizing for this effect, we see that the density of nonself sequences declines monotonically with distance from the nearest self-sequence. Again, this is very well described by our models of nonself sequences (\textbf{Figure 3C-D, Figure S5}), and holds true for parasites, bacteria, and viruses (\textbf{Figure S7C}).\\
\indent The model that we use here does not require that some foreign sequences may have actively evolved to mimic specific human sequence motifs,\cite{maguire2024molecular} but rather a similar distribution might simply stem from shared biophysical constraints for protein folding, metabolism, and function. We do see a small enhancement of exact coincidences in the actual sequences above that predicted by the models, which might reflect evolutionary pressure towards stronger mimicry of genes by pathogens in specific cases.  But even this phenomenon is made easier to evolve by the fact that many pathogen peptides start out only a few mutations away from self.
\subsection*{Immunogenicity as a function of distance from self}

\begin{figure*}[t]
	\begin{center}
		\includegraphics[width=\textwidth]{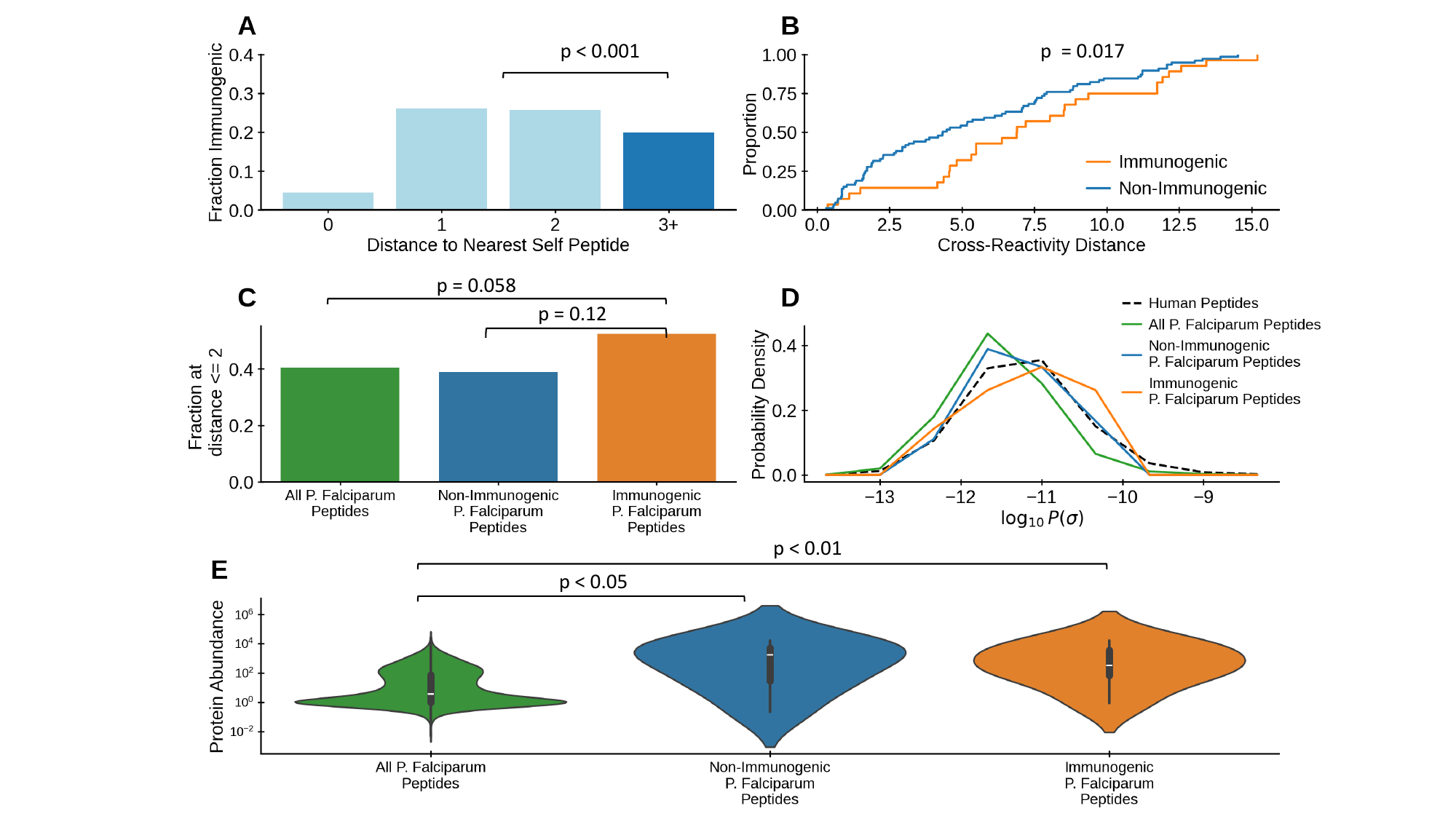}
	\end{center}
		\caption{
		\textbf{Immunogenicity as a function of distance from self} (A) Immunogenicity varies as a function of Hamming distance to the nearest self-peptide from the human reference proteome. Peptides within Hamming distance 2 of the nearest self-peptide were significantly more likely to be immunogenic than peptides at distance 3 or more ($\chi ^2$ test). 
		(B) Empirical cumulative distribution functions (ECDF) of T cell cross-reactivity distances\cite{Luksza2022} for immunogenic (orange) and non-immunogenic (blue) epitopes differing from the nearest self-peptide by a single amino acid. For these distance one peptides, immunogenic epitopes tend towards higher T cell cross-reactivity distance than non-immunogenic epitopes (Kolmogorov–Smirnov test).
		(C) Fraction of different sets of peptides from the \textit{F. Falciparum} proteome that are within Hamming distance 2 of the nearest self-peptide. The background proteome (green) and the non-immunogenic peptides (blue) had smaller fractions of peptides within distance 2 relative to the immunogenic peptides (orange), although these comparisons were not statistically significant with a threshold of 0.05 (green vs. orange, p=0.058 and blue vs. orange, p=0.12, via $\chi ^2$ test).
		(D) Distributions of (log) probabilities of sets of peptides from the F. Falciparum proteome calculated using the maximum entropy model (Eq. 2) for the human proteome.
		(E) Average protein abundance of the source protein of different sets of peptides from the F. Falciparum proteome. Abundance significantly differed between the source proteins for IEDB epitopes (blue and orange) and the background distribution of all peptides (green). Abundance did not differ between immunogenic (orange) and non-immunogenic peptides (blue) (Mann-Whitney U-test).
		Data: Foreign epitopes from IEDB\cite{Vita2019} tested by ELISPOT assay and protein abundance data from Pax-DB\cite{huang2023paxdb} as described in Appendix D. }
\end{figure*}

Our results challenge the notion that the best T cell antigens are far from self.\cite{Calis2013,yarmarkovich2020identification,mohanty2013textbook} This suggests that mutation-derived neoantigens which are close to self are not anomalous, but rather inhabit the same space as pathogenic epitopes. We therefore queried the Immune Epitope Database (IEDB)\cite{Vita2019} to determine the relationship between a peptide’s distance from the nearest self-peptide and its likelihood of being experimentally validated as immunogenic.  In fact, we find that nonself peptides which are one or two amino acids away from self-peptides are more likely to be designated as immunogenic compared to peptides with three or more amino acid substitutions (\textbf{Figure 4A}). As some amino acid substitutions may be more biochemically significant than others, we further characterized the set of peptides that were only one mutation away from self by using a predicted T cell “cross-reactivity distance” between the mutant peptide and the self-peptide, defined in recent work studying tumor neoantigens.\cite{Luksza2022} While some of these hamming-distance-one peptides are biochemically too similar to self and thus presumably rendered not immunogenic by thymic selection, there are many that are sufficiently far away by cross-reactivity distance and can be recognized by the immune system (\textbf{Figure 4B}).\\
\indent Next, we considered epitopes coming from the AT rich genome of \textit{P. falciparum}. While the average \textit{P. Falciparum} peptide comes from a different distribution than a human peptide (\textbf{Figure 2}), there are specific peptides that are disproportionately close to human self-peptides (\textbf{Figure 3B}). Despite the existence of more foreign peptides, immunogenic  \textit{P. Falciparum} epitopes were if anything more likely to be one or two amino acids away from the nearest self-peptide than peptides found not to be immunogenic or than random 9-mers from the background proteome (\textbf{Figure 4C}). In accordance with the high $D_{KL}$, peptides from \textit{P. Falciparum} have a left-shifted (less likely) probability distribution under Eq. (2) which was derived from human proteome statistics (\textbf{Figure 4D}). While non-immunogenic peptides followed this same left-shifted distribution, immunogenic peptides were right-shifted and more closely resemble the human proteome (\textbf{Figure 4D}). Although epitopes annotated in IEDB for \textit{P. Falciparum} tend to come from the most highly abundant proteins, there is not a discernible difference between immunogenic and non-immunogenic epitopes in terms of the abundance of their source proteins (\textbf{Figure 4E}). Thus, despite the wide range of highly foreign epitopes that come from \textit{Plasmodium falciparum}, the immune system surprisingly tends to target the subset of epitopes that are closer and more statistically similar to the self-proteome. 

\section*{DISCUSSION}
\indent Taken together, these results lead to a “shell model” of immunogenicity (\textbf{Figure 5}). Because self and nonself peptides come from the same distribution, and this distribution is strongly nonuniform, an immune system targeting only peptides that are very dissimilar from self is unlikely to find anything to react against. The highest density of nonself targets is in regions of sequence space where there is also the highest density of self-peptides. To maximize the probability of hitting targets and avoiding self, the immune system responds to peptides that are very close to but not exactly self (a shell around self-peptides). Consequently, even when challenged with nonself peptides that come from truly different distributions, like in parasitic infection by \textit{Plasmodium falciparum}, the immune system also targets the subset of epitopes that are closer to self. 

Self/nonself discrimination thus is not a typical classification problem. The usual regime for making distinctions is one in which examples are drawn from very different distributions, as with filtering out spam emails from normal emails. It might be complicated to find the proper axes along which to measure these differences, but they are essential for reliable discrimination. This is not so in self/nonself discrimination (\textbf{Figure 2}); indeed, a good first approximation is that the distributions of peptides found in humans and pathogens are the same.  Discrimination is still possible, but only because self-peptides come from a fixed and finite sample,  $\sim10^{12}$ 9-mers are possible, but only $\sim 10^7$ are realized, whereas the universe of pathogens is effectively unbounded. \\
\begin{figure}[t]
	\begin{center}
		\includegraphics[width=\columnwidth]{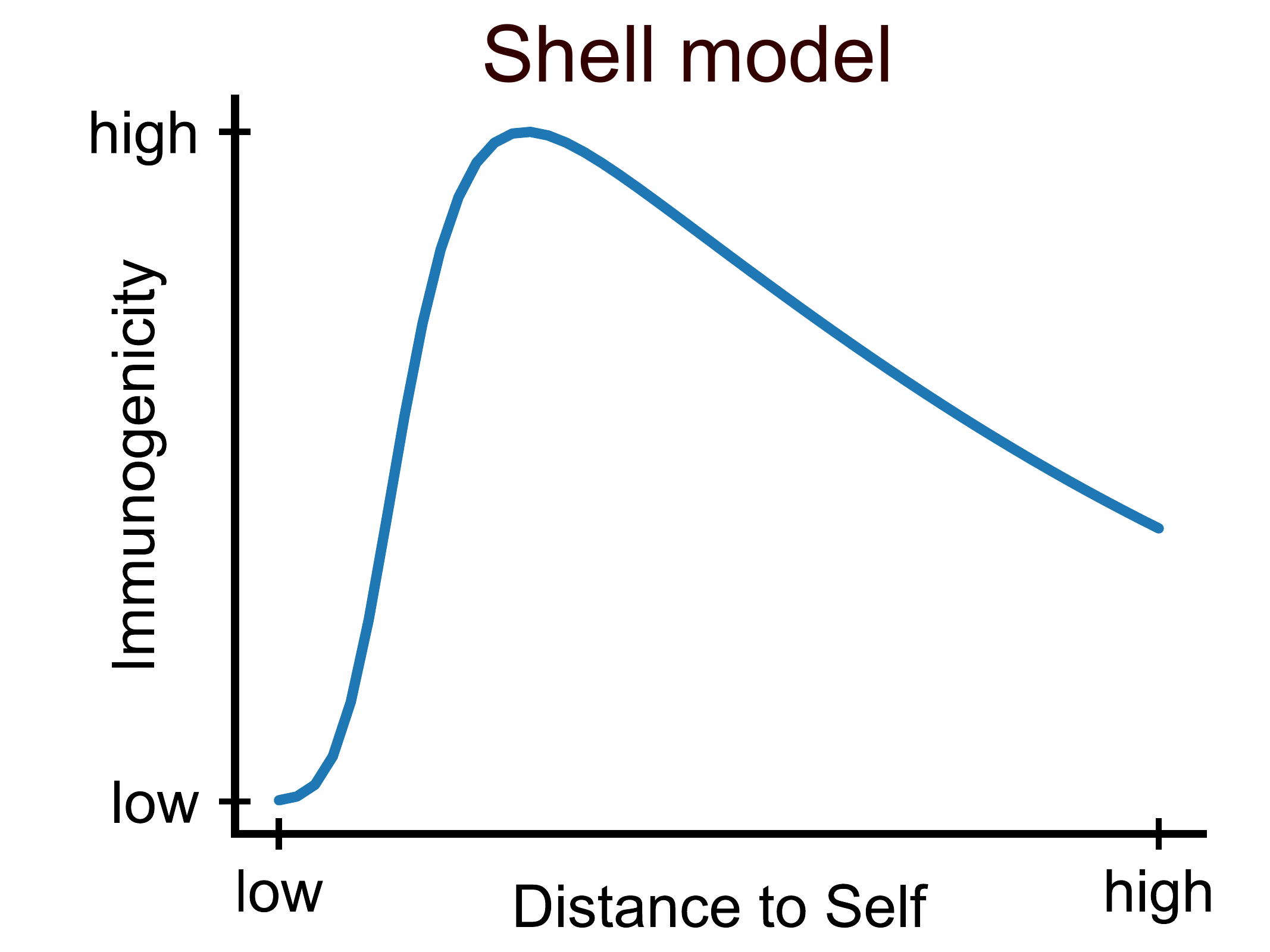}
	\end{center}
	\caption{
		\textbf{Schematic of the predicted biases in immunogenicity.} 
		A shell model of nonself peptide recognition where the immune system biases its search based on the more likely "close to self" peptides. Only (nearly) exact matches to self are non-immunogenic in this model due to tolerance. The most immunogenic peptides are not those that are most foreign, but those in a “shell” around self-peptides.  
	}
\end{figure}
\indent Exploiting the finiteness of the sample usually amounts to “overfitting”,\cite{Mehta2019,Carleo2019} but for the immune system this is essential. For the self/nonself distinction in peptides, overfitting is both allowed and necessary.\cite{George2017} Training on peptides from the self-proteome can guide immune targeting to dense regions of nonself space around self-peptides. This provides a new perspective on the role of positive selection:\cite{Vrisekoop2014,Koncz2021} positively selecting T cells trained on self-peptides would create a set of T cells which can more easily target nonself peptides, especially given T cell cross reactivity. Moreover, by overfitting to the relatively small set of "close to self" peptides, the immune system can leverage a small set of peptides to recognize a pathogen, providing a novel perspective on immunodominance, where indeed a relatively small peptide set drives immune responses. \\
\indent The IEDB measurements and our theory both relate to immunogenicity as an emergent property of the immune system to respond at the collective level and do not explicitly predict the behavior of individual TCRs. Both experimental and theoretical studies\cite{Butler2013,Polonsky2018,polonsky2018induction} have pointed to the role of collective signal integration among T cells in self/nonself discrimination. In light of the current study, an important benefit of such collective decision making is the enabling of discrimination between closely related antigens. The integration of signals from polyclonal T cell pools with a spectrum of avidities to related self and nonself peptides can allow discrimination despite cross reactivity of individual T cells.\cite{kessels2004impact} Additionally, the view of negative thymic selection as "overfitting" also implies the need for peripheral tolerance to deal with false positives, since it is not possible for every T cell to be tested against every possible presented antigen, and self-specific T cells are still present at small numbers in the repertoire.\cite{yu2015clonal} \\
\indent The adaptive immune system has long been viewed as a system for learning the pathogenic environment.\cite{farmer1986immune} T cells are quite capable of, and arguably more likely to, recognize peptides that are close to self. This has important implications for understanding the success of cancer immunotherapies targeting neoantigens, and for guiding epitope selection for vaccines and cellular therapies. If we make assumptions about costs and benefits, we can turn the schematic of \textbf{Figure 5} into quantitative predictions about the optimal parts of the pathogenic environment for the immune system to target.\cite{mayer2015well,mayer2019well} Importantly, the qualitative conclusions of our work are independent of any specific quantitative assumptions: self and nonself are not very different, the distributions of these peptides are strongly inhomogeneous, and the combination of these results means that the immune system must target antigens close to those represented in the organism's own proteome.

\section*{References}
\def\bibsection{}
\bibliographystyle{naturemag}
\bibliography{library}

\section*{ACKNOWLEDGEMENTS}
We thank Gabriel Victora, Vinod Balachandran, Chrysothemis Brown, Curtis Callan, Warren James, Marta Luksza, Taha Merghoub, Andrea Schietinger, and Zachary Sethna for insightful discussions.  This work was supported in part by the National Science Foundation through the Center for the Physics of Biological Function (PHY--1734030); by a Lewis--Sigler fellowship (AM); by a Medecine Sciences fellowship of the Fondation pour la Recherche Medicale (QM); by fellowships from the Simons Foundation and the John Simon Guggenheim Memorial Foundation (WB); and by the Lustgarten Foundation, the Mark Foundation (ASPIRE Award), the NCI (P30CA008748, U01CA228963), the Pershing Square Sohn Foundation (Pershing Square Sohn Prize–Mark Foundation Fellowship), and Stand Up to Cancer (BDG). 
 
\section*{DECLARATION OF INTERESTS}
B.G. has received honoraria for speaking engagements from Merck, Bristol Meyers Squibb, and Chugai Pharmaceuticals; has received research funding from Bristol Meyers Squibb and Merck; and has been a compensated consultant for Darwin Health, Merck, PMV Pharma and Rome Therapeutics of which he is a co-founder. All other authors declare no conflict of interests.

\section*{DATA AVAILABILITY STATEMENT}

Original data will be made available upon reasonable request.
\section*{CODE AVAILABILITY STATEMENT}
All code is available on github at \url{https://github.com/andim/peptidome}

\section*{Appendix A: Human and pathogen sequences}
\label{App:data}

We compared proteome statistics from model organisms across different branches of jawed vertebrates: human and mouse as two examples of mammals, chicken as an example of a bird, and zebrafish as an example of a jawed fish. We also compared them against a set of pathogen proteomes: a collection of human viruses (individual viral proteomes are too small for reliable statistical analyses), several pathogenic bacterial species ({\em Mycobacterium tuberculosis, Listeria monocytogenes, and Streptococcus pyogenes}), as well as a parasite ({\em Plasmodium falciparum}). These examples were chosen because of their representation in the database of immune responses (See section \ref{App:iedb}). 

\begin{table*}[t!]
	\begin{tabular}{cccc}
		Short name&Full name&Proteome ID& Approximate Length  \\ \hline
		Human&Homo sapiens&UP000005640& \num{1e7} \\
		Mouse&Mus Musculus&UP000000589& \num{1e7} \\
		Chicken&Gallus gallus&UP000000539& \num{1e7} \\
		Zebrafish&Danio rerio&UP000000437 & \num{1e7} \\
		Malaria&Plasmodium falciparum&UP000001450& \num{4e6} \\
		Tuberculosis&Mycobacterium tuberculosis&UP000001584& \num{1e6} \\
		Listeria&Listeria monocytogenes&UP000000817& \num{8e5} \\
		StrepA&Streptococcus pyogenes&UP000000750& \num{5e5} \\
		Human Virome&Human Viruses&Uniref90-Filtered&\num{1e6}
	\end{tabular}
	\caption{Reference proteomes used in this study.\label{datalist}}
\end{table*}

For each species we downloaded its reference proteome from Uniprot \cite{UniprotConsortium2021} using the proteome identifiers shown in \textbf{Table \ref{datalist}}. The table also displays shortened names used in main text figure legends. For the pan-viral proteome we downloaded all viral sequences annotated with human as a host species (taxon id: 9606) from Uniref90, a protein database clustered at 90.  For each proteome we then generated a list of all possible 9-mers by iterating over all possible starting positions within each protein from the proteome. We note that the total number of peptides is roughy equal to the total length of the proteome (except for forbidden start positions at the edges of the protein).

\section*{Appendix B: Maximum entropy models}
\subsection*{Model Definition}
\label{App:maxent}

We use the maximum entropy framework \cite{Jaynes1955} as a principled way to include increasingly detailed statistical structure into a series of nested models for peptide statistics. Using this approach we constrain average features of the sequence $\langle f_\mu(\boldsymbol \sigma)\rangle$ to equal their empirical values $\bar{f_\mu}$, while otherwise keeping the probability distribution as random as possible. Mathematically, this means we choose a probability distribution that maximizes the Shannon entropy
\begin{equation}
	S[P(\B \sigma)] = - \sum_{\B \sigma} P(\B \sigma) \log P(\B \sigma),
\end{equation}
subject to the normalization constraint $\sum_{\B \sigma} P(\B \sigma) = 1$, and constraints that enforce the equality of modelled and empirical observables
\begin{equation}
	\langle f_\mu(\boldsymbol \sigma)\rangle = \sum_{\boldsymbol \sigma} P(\boldsymbol \sigma) f_\mu(\boldsymbol \sigma) = \bar{f_\mu}.
\end{equation}
Optimizing with respect to the normalization constraint yields a Boltzmann distribution of the form,
\begin{equation}
	P(\boldsymbol \sigma) = \frac{1}{Z} \exp\left[ -E(\B \sigma) \right],
\end{equation}
where
\begin{equation}
	E(\B \sigma) = \sum_{\mu=1}^K \lambda_\mu f_\mu(\boldsymbol \sigma),
\end{equation}
is a sum of terms involving each constraint, and 
\begin{equation}
	Z = \sum_{\B \sigma} \exp \left[ - E(\B \sigma) \right]
\end{equation}
is a normalization factor, called the partition function.  This is mathematically equivalent to the statistical mechanics of a system in thermal equilibrium, where $ E(\B \sigma)$ is the energy as a function of its configuration.

To fix the parameters $\lambda_\mu$, we follow a standard method: At current values of the parameters we  estimate $\langle f_\mu(\B \sigma)\rangle$ using Monte Carlo sampling,  then change parameters to reduce the discrepancy between estimated and empirical observables, and iterate until these discrepancies are nearly zero. \cite{Ackley1985}

Correlations extend beyond the scale of peptides of interest (\textbf{Figure S1}), so we start by considering compositional constraints on the covariation of the total count of amino acids of different types. That is, we 
count the number of amino acids of each type in the peptide,
\begin{equation}
	n^\alpha(\B \sigma) = \sum_{\rm i} s_{\rm i}^\alpha, 
\end{equation}
and constrain its expectation value, and then do the same with the second and third moments
\begin{align}
	n^{\alpha\beta}(\B \sigma) &= n^\alpha(\B \sigma) n^\beta(\B\sigma) = \left(\sum_{{\rm i}=1}^L s_{\rm i}^\alpha\right) \left(\sum_{{\rm j}=1}^L s_{\rm j}^\beta\right) 
\end{align}
\begin{align}
	n^{\alpha\beta\gamma}(\B \sigma) &= n^\alpha(\B \sigma) n^\beta(\B\sigma) n^\gamma(\B\sigma) \nonumber\\
	&= \left(\sum_{{\rm i}=1}^L s_{\rm i}^\alpha\right) \left(\sum_{{\rm j}=1}^L s_{\rm j}^\beta\right) \left(\sum_{{\rm k}=1}^L s_{\rm k}^\gamma\right).
\end{align}
This leads to a maximum entropy probability distribution with an effective energy
\begin{equation}
	E(\boldsymbol \sigma) = \sum_{\alpha=1}^{20} \lambda_1^\alpha n^\alpha +  \sum_{\alpha,\beta=1}^{20} \lambda_2^{\alpha\beta}n^\alpha n^\beta + \sum_{\alpha,\beta,\gamma=1}^{20}  \lambda_3^{\alpha\beta\gamma}  n^\alpha n^\beta n^\gamma .
\end{equation}
This model only involves global couplings between amino acids independent of their distance.

Another common constraint involves the two-point frequencies $ f_{\rm ij}^{\alpha\beta}(\B \sigma) = s_{\rm i}^\alpha s_{\rm j}^\beta$; the expectation value $\langle f_{\rm ij}^{\alpha\beta}\rangle$ is the probability of finding amino acid of type $\alpha$ at site $\rm i$ and amino acid of type $\beta$ at site $\rm j$. By construction the data are translation invariant, except for edge effects arising from the finite length of proteins, so this should depend only on the distance $\Delta =|{\rm i} - {\rm j}|$ between the two amino acids. We explicitly enforce this  invariance in the model by instead constraining the expectation value of 
\begin{align}
	f_{\Delta}^{\alpha\beta}(\B \sigma) &= {1\over 2} \sum_{\substack{{\rm i,j}=1\\ {|\rm i}-{\rm j}|=\Delta}}^9 s_{\rm i}^\alpha s_{\rm j}^\beta.
\end{align}

Taking all terms together, we obtain the energy of the full model (with $J_0^{\alpha\beta} = 0$)
\begin{widetext}
	\begin{equation}
		E({\boldsymbol \sigma}) = \sum_{\alpha=1}^{20} \lambda_1^\alpha n^\alpha + \sum_{\alpha,\beta=1}^{20} \lambda_2^{\alpha\beta} n^\alpha n^\beta + \sum_{\alpha,\beta,\gamma=1}^{20} \lambda_3^{\alpha\beta\gamma} n^\alpha n^\beta n^\gamma + {1\over 2}\sum_{{\rm i},{\rm j} = 1}^9 \sum_{\alpha ,\beta=1}^{20} J^{\alpha\beta}_{\Delta}  s_{\rm i}^\alpha s_{\rm j}^\beta, \label{model2} 
	\end{equation}
\end{widetext}
All of the coefficients must be adjusted so that the predictions of the model match observed features of the data: $\lambda_1^\alpha$ is fixed by matching the mean number of amino acids of type $\alpha$; $\lambda_2^{\alpha\beta}$ by matching the covariance in numbers of amino acids of types $\alpha$ and $\beta$; $\lambda_3^{\alpha\beta\gamma}$ by matching the third moments of these numbers;
the matrix $J^{\alpha\beta}_{\Delta}$, where $\Delta=|{\rm i}-{\rm j}|$, must be adjusted to match the distance dependence of the pairwise correlations; and the constant (partition function) $Z$ serves to normalize the distribution, so that
\begin{equation}
	Z = \sum_{\boldsymbol\sigma}  \exp\left[ - E({\boldsymbol \sigma})\right] .
\end{equation}

To test the predictive power of the maximum entropy model, we compare the density of states between model and data, and we calculate triplet correlations according to
\begin{equation}
	C_{\rm ijk}^{\alpha\beta\gamma} = \langle \delta s_{\rm i}^\alpha \delta s_{\rm j}^\beta \delta s_{\rm k}^\gamma \rangle ,
	\label{triplet}
\end{equation}
with $\delta s_{\rm i}^\alpha =  s_{\rm i}^\alpha - \langle  s_{\rm i}^\alpha\rangle$.

\subsection*{Calculating entropies and statistical divergences using thermodynamic integration}
\label{App:thermo_int}

How can we determine the entropy of a fitted model? From the definition of the model in terms of a Boltzmann distribution,
\begin{equation}
	P(\B \sigma) = \frac{1}{Z}\exp[-E(\B\sigma)],
\end{equation}
we obtain the entropy
\begin{align}
	S[P(\B \sigma)] &\equiv - \sum_{\B \sigma}  P(\B \sigma) \log P(\B \sigma)  \\
	&= \langle E(\B \sigma) \rangle_{P(\B\sigma)} - F,
\end{align}
where $F = - \log Z$.
Similarly, we can express the Kullback-Leibler divergence as
\begin{align}
	D_{KL}[P(\B \sigma) || Q(\B \sigma)] &= - \sum_{\B \sigma}  P(\B \sigma) \log \frac{P(\B \sigma)}{Q(\B \sigma)},  \\
	&= \langle \Delta E(\B \sigma) \rangle_{P(\B \sigma)} - \Delta F,
\end{align}
where $\Delta E = E_Q(\B \sigma) - E_P(\B \sigma)$ and $\Delta F = F_Q - F_P$.
In both instances, we can approximate the expectation value over $P(\B \sigma)$ as the mean over Monte Carlo samples drawn from that distribution.
But we also need to determine $F$, or  the partition function of the fitted models.

As an exact evaluation of the partition function is computationally intractable we use thermodynamic integration \cite{Marchi2019b} to numerically approximate $F = -\log Z$:
We define the perturbed energy function
\begin{equation}
	E_g (\B \sigma) = E_{\rm ref}(\B \sigma) + g \Delta E(\B\sigma),
\end{equation}
where $E_{\rm ref}(\B \sigma)$ is a reference energy for which $F_{\rm ref}$ can be calculated analytically, and $0 \leq g  \leq 1$ is a parameter scaling the additional energy term. 
For the maximum entropy model defined in Eq (\ref{model2}) we use
\begin{align}
	E_{\rm ref} =& \sum_{\alpha =1}^{20} \lambda_1^\alpha n^\alpha,   \\
	\Delta E({\boldsymbol \sigma}) =& \sum_{\alpha,\beta=1}^{20} \lambda_2^{\alpha\beta} n^\alpha n^\beta + \sum_{\alpha,\beta,\gamma=1}^{20} \lambda_3^{\alpha\beta\gamma} n^\alpha n^\beta n^\gamma \nonumber\\
	&\,\,\,\,\,\,\,\,\,\, + {1\over 2}\sum_{{\rm i},{\rm j} = 1}^9 \sum_{\alpha ,\beta=1}^{20} J^{\alpha\beta}_{\Delta}  s_{\rm i}^\alpha s_{\rm j}^\beta.
\end{align}
Notice that $g =1$, $E_g (\B \sigma) = E (\B \sigma)$ in  Eq (\ref{model2}).

We have chosen  $E_{\rm ref}$ to describe a model in which amino acids are chosen independently at each site, and we can see that  
\begin{equation}
	F_{\rm ref} = -k \log \left[ \sum_{\alpha=1}^{20} \exp\left(\lambda_1^\alpha\right)\right] .
\end{equation}
In contrast,  $\Delta E$ includes all couplings between different residues in the $k$--mer. Thus as we move along the family of models from $g = 0$ to $g =1$, we interpolate between the  independent model and the true model.

We can define the free energy $F_g$ for the model at a specified value of the parameter $g$. Importantly, we have
\begin{equation} \label{eqdFalpha}
	\frac{\ud F}{\ud g}  = - \langle \Delta E(\B \sigma) \rangle_{P_g (\B \sigma)},
\end{equation}
where we emphasize again that the right hand side can be approximated as the mean over Monte Carlo samples.  
Thus we can calculate the free energy of the real model at $g=1$ using ``thermodynamic integration''
\begin{equation}
	F(g=1) = \int_0^1 \ud g \frac{\ud F}{\ud g}  + F_{\rm ref}
\end{equation}
In practice, we draw Monte Carlo samples from the perturbed models with evenly spaced $g \in [0, 1]$, use these samples to evaluate ${\ud F}/{\ud g}$, and  then evaluate the integral by Simpson's rule.

\subsection*{Model Sampling}
Samples were drawn from the model using the Markov Chain Monte Carlo (MCMC) based Metropolis-Hastings Algorithm. \cite{hastings1970monte} Samples for the original models were drawn using a chain length of $5418757$ and thinning parameter of $60$. Samples for the filtered models with MHC were drawn using a longer chain length of $1e7$ (so that enough samples remained after the filter) and thinning parameter of $60$. The estimates calculated via sampling did not depend significantly on the choice of hyperparameters (\textbf{Figure S9}).

\subsection*{Effective Diversity}
Effective Diversity is calculated as $D=e^S/20^9$, where S is the estimated  Shannon entropy (see above \ref{App:thermo_int}) and D is the Shannon diversity (also known as perplexity) relative to that of a uniform distribution over 9-mers.

\subsection*{Coincidence Probability}
Coincidence probability was estimated using the Simpson's index \cite{Simpson1949,Hunter1988} on samples drawn from the models:
\begin{equation}
	D = \sum_{i=1}^S \frac{n_i (n_i - 1)}{N (N - 1)}
\end{equation}
where $N$ is the total number of peptides in the sample population, S is the number of unique peptides in the sample, and $n_i$ is the number of occurences of the $ith$ peptide in the sample.

\section*{Appendix C: Combined MHC and Maximum Entropy Distribution}
To model the effects of MHC presentation, we introduce a filtered probability distribution for 9-mers in a proteome that are presented on an HLA haplotype $H = \{\ A_1, A_2, B_1, B_2, C_1, C_2  \}$. We assume that presentation and maximum entropy probability are independent, and thus:

\begin{equation}
	{\widetilde{P}}(\B \sigma) = P(\B \sigma) Q(\B \sigma, H)
\end{equation}
where $P(\B \sigma)$ is the fit maximum entropy model, and $Q(\B \sigma, H) $ describes the probability that peptide $\B \sigma$ is presented on at least one allele in a haplotype $H$. For each peptide $\B \sigma$ and HLA allele $\B h$, we use netMHC-4.0 \cite{Andreatta2016} to get a rank of predicted binding affinity for $\B \sigma$ on  allele $h$.  We define the peptide's presentability score as:
\begin{equation} 
	s(\B \sigma, H) = min \{ Rank (\B \sigma, h);  h \in H \}
\end{equation}
We design $Q(\B \sigma, H)$ as a binary filter, only keeping peptides that are a strong binder to at least one of the alleles in the haplotype. We use the \% Rank cutoff of 0.5\% as advised by the authors of NetMHC-4.0:

\begin{equation} 
	Q(\B \sigma, H) \propto   \begin{cases} 1 & s < 0.5\% \\ 0  & s>= 0.5\% \\	\end{cases} 
\end{equation}.

To estimate the entropy of the filtered model, we sample from the maximum entropy distribution and filter the sample using netMHC to only keep strong binders. The entropy is then calculated from the filtered, sampled set as:
\begin{align*} 
	S[{\widetilde{P}}(\B \sigma)] &= -\sum_{\B \sigma} \widetilde{P}(\B \sigma)\log\widetilde{P}(\B \sigma)
	\\
	&= -\sum_{\B \sigma}\widetilde{P}(\B \sigma)(\log Q(\B \sigma) + \log P(\B \sigma))
	\\
	&=  -\sum_{\sigma}\widetilde{P}(\sigma)\log P(\sigma) -\sum_{\sigma}\widetilde{P}(\sigma)\log Q(\sigma)
	\\
	&=   -F + \langle E(\sigma) \rangle_{\widetilde{P}}  - \log \frac{1}{b} 
\end{align*}
where $b$ is the fraction of the sample that survived the binding filter ($b=\frac{n_{binders}}{N}$), and $ \langle E(\sigma) \rangle_{\widetilde{P}}$ is a Monte Carlo estimate of the cross-entropy term relating the new distribution to the original model distribution.  
Intuitively, this means that a more selective filter (smaller b) leads to less entropy in the filtered distribution, and a filter that keeps $100\%$ of the entire orginal sample does not change the entropy at all $(\log 1 = 0)$.

\subsection*{MHC Haplotypes}
In order to characterize canonical human haplotypes, we used previous work \cite{hoyos2022fundamental} that characterized a haplotype multinomial distribution using data from  from the National Marrow Donor Program (NMDP) database, which provides information on 1,242,890 donors with full MHC-I linkage information.\cite{gonzalez2020allele, gragert2013six} Following this previous work, \cite{hoyos2022fundamental} we assume the MHC-I haplotype will consist of two each of HLA-A, HLA-B, and HLA-C for a total of 6 MHC-I genes. We used the previously derived multinomial distribution with an independent frequency model without MHC-I linkage for each HLA-A, HLA-B, and HLA-C gene. As described in the previous work \cite{hoyos2022fundamental}, the number of heterozygous HLA-I genes is given by $N_H$ , where $N_H \in [0,1,2,3]$. The probability of a haplotype $H=[A_1,A_2,B_1,B_2,C_1,C_2]$ is given by:
\begin{equation}
	p(H) = 2^{N_H} \prod_{h \in H} p_h \label{eq:hap}
\end{equation}
where $p_h$ corresponds to the marginal probability of HLA-I $h$ within haplotype $H$. A visualization of this distribution can be seen in \textbf{Figure S2}. For many of the analyses in this paper, we used the top 500 most probable haplotypes as determined by Equation~\ref{eq:hap}.

\section*{Appendix D: DISTANCE DISTRIBUTIONS AND IMMUNOGENIC EPITOPES}
\subsection*{Hamming Distance Distributions}
\label{App:nn}
We  aligned peptides to their nearest self-peptide by Hamming distance using an exact algorithmic approach,\cite{chotisorayuth2024lightning} leveraging publically available code in pyrepseq \url{https://github.com/andim/pyrepseq}. We note that owing to the size of the proteomes (\textbf{Table \ref*{App:data}}), estimating the error on these distributions using normal approximation gives errors that are basically zero. For example, the length of the p. Falciparum proteome is on the order of $10^6$, and a simple calculation of $\sqrt{(p*(1-p)/N)}$ gives standard errors too small to show up ($\approx 10^{-4}$)

\subsection*{Epitope Immunogenicity}
\label{App:iedb}
To understand how the immunogenicity of foreign epitopes depends on their distance to the nearest self peptides we analyzed data from the Immune Epitope Database (IEDB). \cite{Vita2019} This database contains a collection of experimental data about T cell activation in response to different peptides. The experiments recorded in the database involve assays, such as the enzyme-linked immunosorbent spot (ELISPOT) assay, in which T cells from a donor sample are cultured in the presence of an antigen. This allows assessing the reactivity of antigen-specific T cells by counting cells that activate in response to the antigen and secrete cytokines. The ability of an epitope to elicit an immune response as measured by such an assay is commonly called the immunogenicity of the epitope.

\begin{table}[b]
	\begin{tabular}{ccc}
		Distance | &Count | &Immunogenic Count \\ \hline
		0&22&1 \\
		1&107&28 \\
		2&2421&622 \\
		3&3496&694 \\
		4+&14&6 \\
	\end{tabular}
	\caption{IEDB T Cell Epitopes \label{iedblist}}
\end{table}

To perform our analyses we downloaded the full T cell reactivity dataset from IEDB (\url{http://www.iedb.org/ downloader.php?file_name=doc/tcell_full_v3.zip}) on 2022/07/05. We extracted all epitopes presented on human MHC class I, with a length of 9 amino acids and not annotated as coming from the host. We excluded epitopes of foreign origin tested in the context of autoimmune disease, allergy, and cancer, as these might introduce a sampling bias towards “special” peptides close to self but immunogenic. We define epitopes as immunogenic, if their recorded qualitative score was “Positive,” “Positive-Low,” or “Positive-High”. In cases, for which multiple independent measurementes where available we only retained epitopes with concordant immunogenicity assessment across studies.

We then aligned foreign epitopes to their nearest self-peptide by Hamming distance as described above (Section \ref{App:nn}).  The number of peptides and number of those that are immunogenic for each distance bin can be found in \textbf{Table \ref{iedblist}}.

We performed  a $\chi^2$-test between peptides that are  distance <= 2 and peptides that are distance >2 as implemented in the Python package statsmodels v0.12.2. For \textbf{Figure 4A}, $651/2550=25.5\%$ of the "close" peptides were immunogenic, compared to only $700/3515=19.9\%$ of the "far" peptides. This revealed a statistically significant drop in immunogenicity for far peptides (at Hamming distance 3+) compared to close ones at distance 2 or less. This suggests the immune system naturally recognizes foreign epitopes close to self, even when farther peptides, more dissimilar from self, are available to be recognized. The same qualitative finding is also supported by a recently published study, \cite{koncz2021self} in which the analysis of epitope sequence similarity was restricted to T cell exposed motifs.
IEDB reports results from multiple different experimental techniques and we found evidence that data might not be comparable across experimental approaches. Specifically, we found that assays differed greatly in the fraction of reported immunogenic epitopes (\textbf{Figure S8A}). The analysis reported in \textbf{Figure 4} report results from a single assay, the ELISPOT assay, which is the most commonly reported within the database. An analysis of data from other assays shows less clear signal (\textbf{Figure S8B-F}).
Interestingly, we find that epitopes with exact matches in the self-proteome are largely non-immunogenic when assessed by ELISPOT, but find multiple instances of such epitopes annotated as immunogenic for other assays. This is in line with experimental evidence that ELISPOT assays are uniquely sensitive to peripheral tolerance mechanisms, as supported by increased responses following depletion of regulatory T cells. \cite{Bonertz2009} We also find that low T cell cross-reactivity distance of epitopes with single amino acid substitutions makes epitopes more likely (although not statistically significant) to be assessed as immunogenic by non-ELISPOT assays (\textbf{Figure S8C}), in contrast to the expected, reverse trend observed in ELISPOT data (\textbf{Figure 4B}).

\subsection*{Plasmodium Falciparum Epitopes}
While the distributions of vertebrate and viral proteomes are largely similar to the human proteome (\textbf{Figure 2A}), the proteome of the malaria parasite \textit{Plasmodium falciparum}, an organism known to have an unusually AT-rich genome, \cite{Hamilton2017} has a larger divergence. To understand how the immunogenicity of foreign epitopes from this parasite depends on their distance to the nearest self peptides we analyzed data from IEDB annotated as being from human host with 'Parent Species' = 'Plasmodium falciparum'. We filtered to MHC Class-I epitopes as above, and additionally for any MHC annotation that was missing in the database, we manually looked at the original reference and kept the epitope if the original contributing literature specified that the epitope was for a class-I HLA.  This resulted in 151 positive epitopes, 60 of which were based on ELISPOT. 

\subsection*{Plasmodium Falciparum Protein Abundance}
Protein abundance data was downloaded from Pax-DB\cite{huang2023paxdb} at \url{https://pax-db.org/dataset/5833/3599693021/}. This data was mapped to IEDB data using the Uniprot identifier map \url{https://pax-db.org/downloads/5.0/paxdb-uniprot-links-v5.0.zip}.

\subsection*{Cross-Reactivity Distance}
For hamming-distance-one epitopes, we calculated cross-reactivity distance as previously. \cite{Luksza2022} Publically available code to compute this metric can be found at \url{https://github.com/LukszaLab/NeoantigenEditing}. Cross-reactivity distances were compared between Immuognenic and Non-Immunogenic hamming-distance-one epitopes using their emprical cumulative distributions and significance was tested using a two-sample Kolmogorov-Smirnov as implemented in scipy's \texttt{ks\_2samp} function.

\appendix
\renewcommand{\thefigure}{S\arabic{figure}}
\setcounter{figure}{0}  

\begin{figure*}[!]
	\begin{center}
		\includegraphics[width=.9\textwidth]{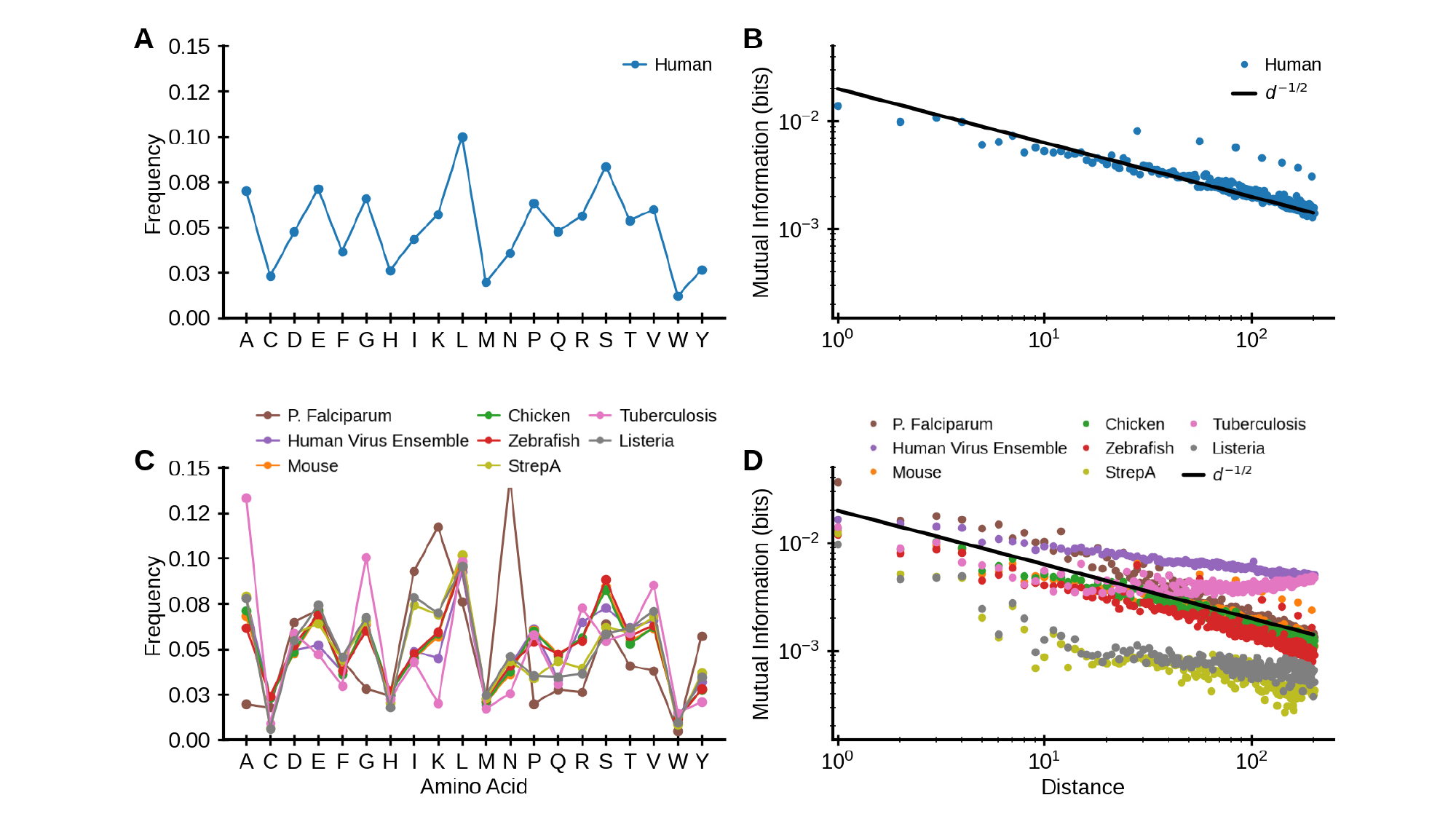}
	\end{center}
	\caption{
		\textbf{Proteome statistics}
		Basic statistics for the human reference proteome (A-B) and various other proteomes (C-D) analyzed in this paper. For details of the databases used, see Methods.
		(A): Amino acids are used with unequal probabilities in the human proteome. 
		(B): Correlations between amino acids extend across a very long distance. Decay of mutual information with separation for the human proteome. Here the mutual information in bits per symbol is shown as a function of separation distance $d(X,Y) = |\rm i - j|$, where the symbols X and Y are located at positions i and j in the protein sequence. This decay is modeled (black) as a power law\cite{lin2016criticality} with exponent $-1⁄2$.
		(C): Same as (A) but for several example non-human vertebrate proteomes and various pathogen proteomes.		(D): Same as (B) but for several example non-human vertebrate proteomes and various pathogen proteomes}
\end{figure*}

\begin{figure*}[t!]
	\begin{center}
		\includegraphics[width=\textwidth]{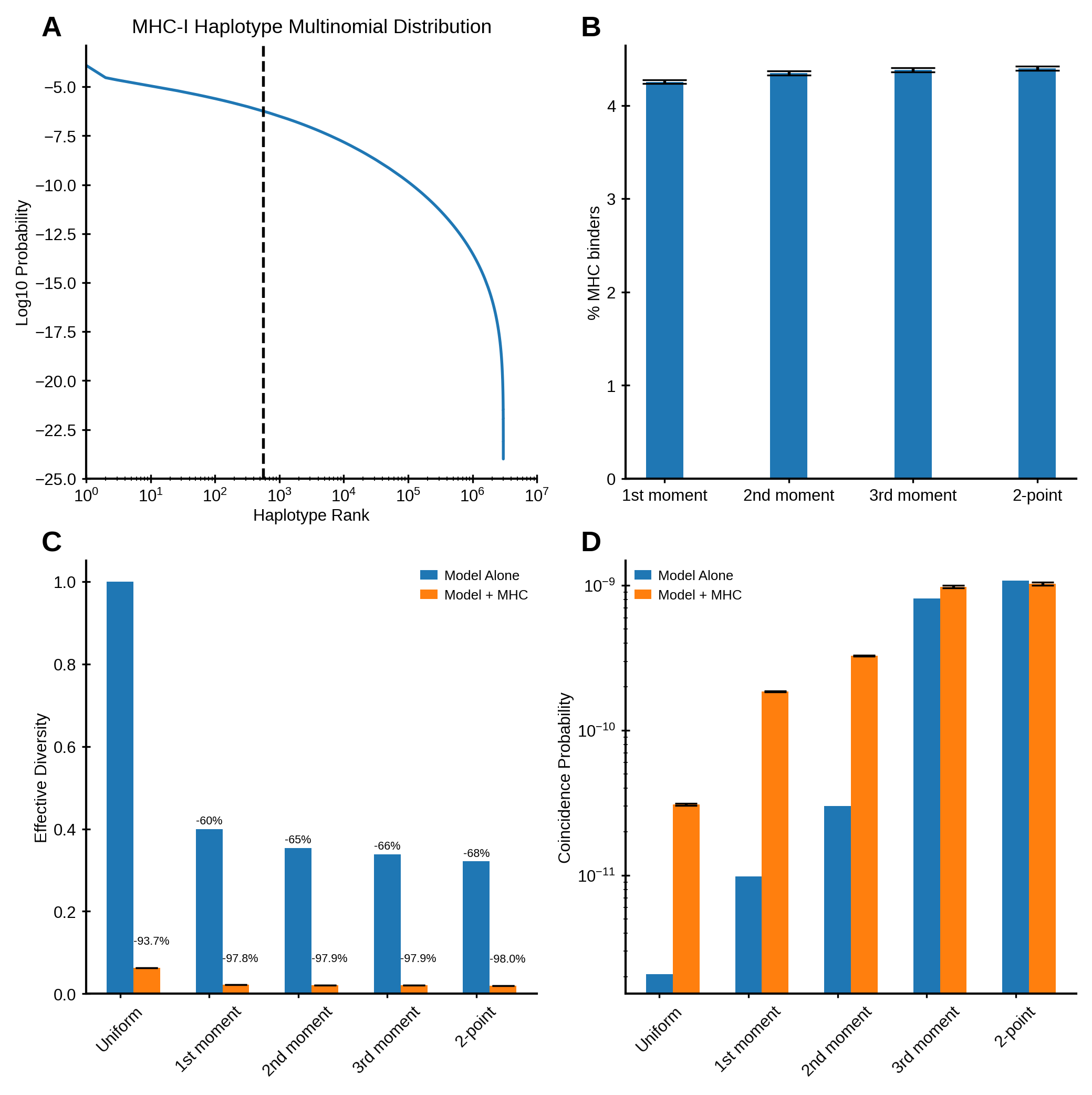}
	\end{center}
	\caption{
		\textbf{Joint maximum entropy and MHC models}
		(A)	Ranked probability distribution of MHC-I haplotypes, as described in Appendix C. Haplotypes that ranked within the top 500 most probable were used in this analysis (cutoff shown as dotted black line). 
		(B)	Percentage of peptides in samples from each model that were predicted by netMHC-4.0\cite{Andreatta2016} to be strong binders to at least one allele in the haplotype. Error bars show SEM across the top 500 haplotypes.
		(C)	Effective Diversity for each model alone (blue) and each model joint with MHC-I presentation (orange). Error bars for joint models (orange) shown SEM across the top 500 haplotypes.
		(D)	Probability of coincidence for each model alone (blue) and each model joint with MHC-I presentation (orange). Error bars for joint models (orange) shown SEM across the top 500 haplotypes.}
\end{figure*}

\begin{figure*}[t!]
	\begin{center}
		\includegraphics[width=\textwidth]{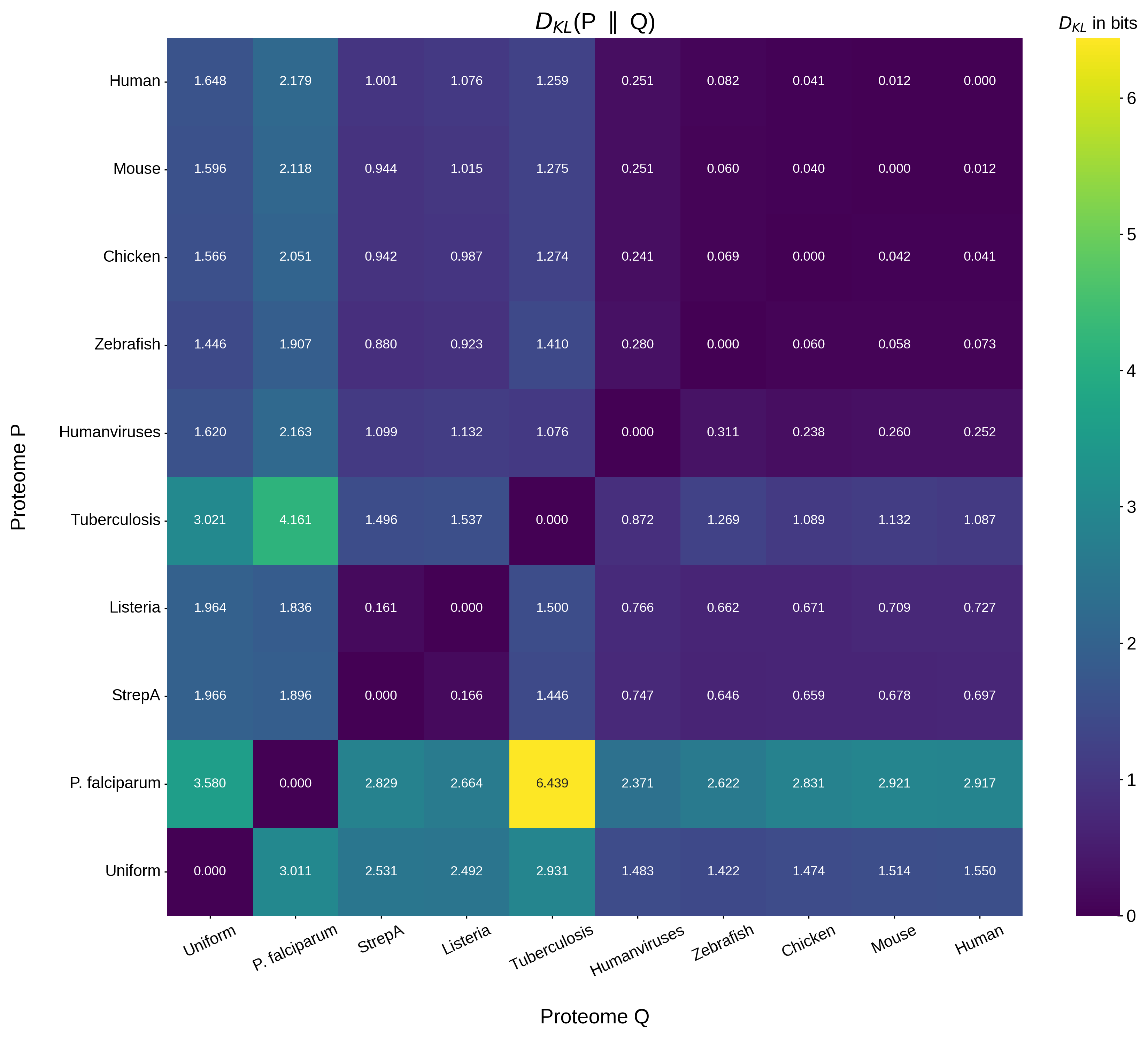}
	\end{center}
	\caption{
		\textbf{Pairwise divergence between peptides from different proteomes}
Pairwise Kullback-Leibler (KL) divergences between peptide distributions of different proteomes. For each proteome we show the statistical distance calculated according to the full 2-point model with all of the constraints. The first column and the last column here correspond to the y and x coordinates respectively for the diamond shaped points in \textbf{Figure 2}.}

\end{figure*}

\begin{figure*}[t!]
	\begin{center}
		\includegraphics[width=\textwidth]{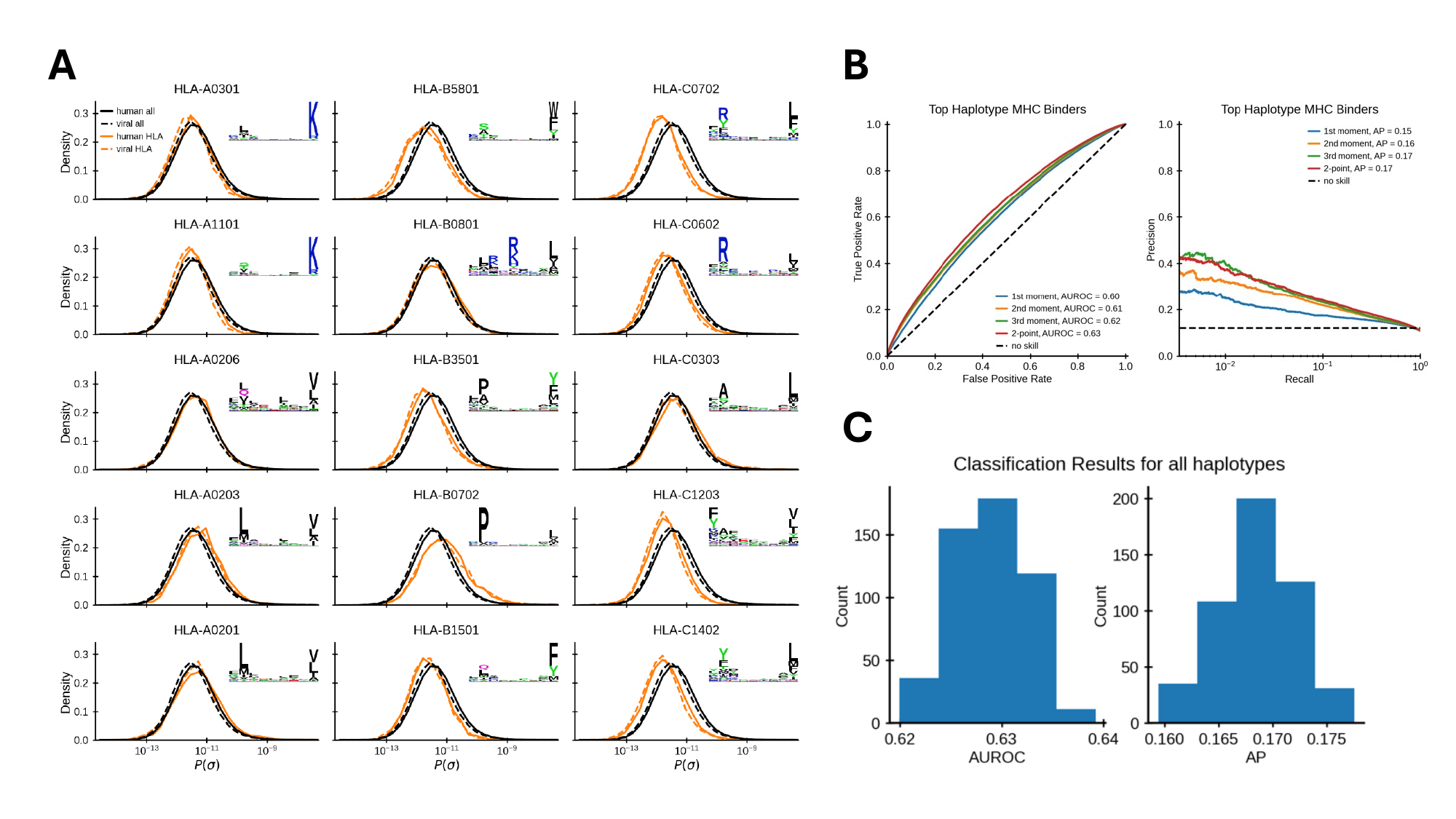}
	\end{center}
	\caption{
		\textbf{Classification of HLA restricted peptides}
		(A)	Each subplot displays the distribution of model probabilities for HLA restricted peptides predicted by netMHC-4.0\cite{Andreatta2016} for the top 5 HLA-A/B/C alleles (in terms of number of training sequences). The model probabilities are predicted by Eq. (2) with parameters inferred from the human proteome. For each HLA allele, the plot displays the distributions both of predicted self-peptide binders (solid orange lines) and predicted binders from human viruses (dashed orange lines). Each plot also shows the distributions of all peptides regardless of HLA restriction for reference (black solid/dashed lines), alongside a sequence logo of all predicted binders that highlights which residues are enriched in anchor positions.
		(B)	ROC and PR Curves for classifier using model energies (as in Fig 2B-C) for restricted test set of peptides that bind to an allele in the most common HLA haplotype (rank=1 in Fig S2A).
		(C)	Distribution of AUROC and AP values for same classification scheme, across each of the top 500 ranking haplotypes.}
	\end{figure*}
	
	\begin{figure*}[t!]
		\begin{center}
			\includegraphics[width=\textwidth]{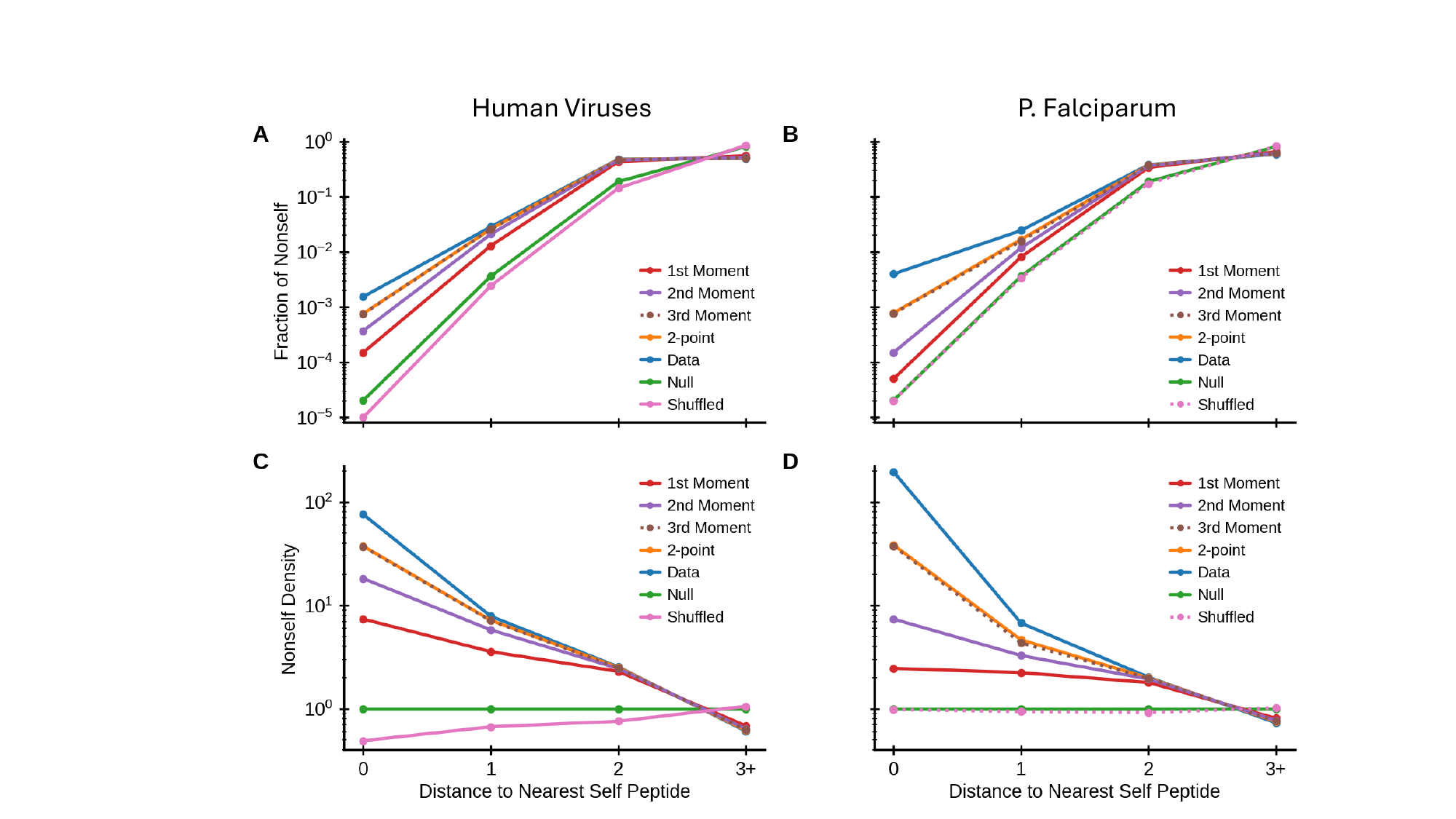}
		\end{center}
		\caption{
			\textbf{Distance to nearest self-peptide as a function of model constraints}
			(A-B) Distribution of distances to the nearest self-peptide for peptides from human viruses (A) and P. Falciparum (B) as found in the data (in blue, see Methods), in predictions from the family of models (red, purple, brown, orange), in a null model with a uniform distribution over all 209 9–mers (green), and model with non-uniform but shuffled probabilities for amino acid usage (pink). (C-D) Relative density of distances to the nearest self-peptide for the same sets of peptides. Colors as in (A-B).}
	\end{figure*}
	\begin{figure*}[t!]
	\begin{center}
		\includegraphics[width=\textwidth]{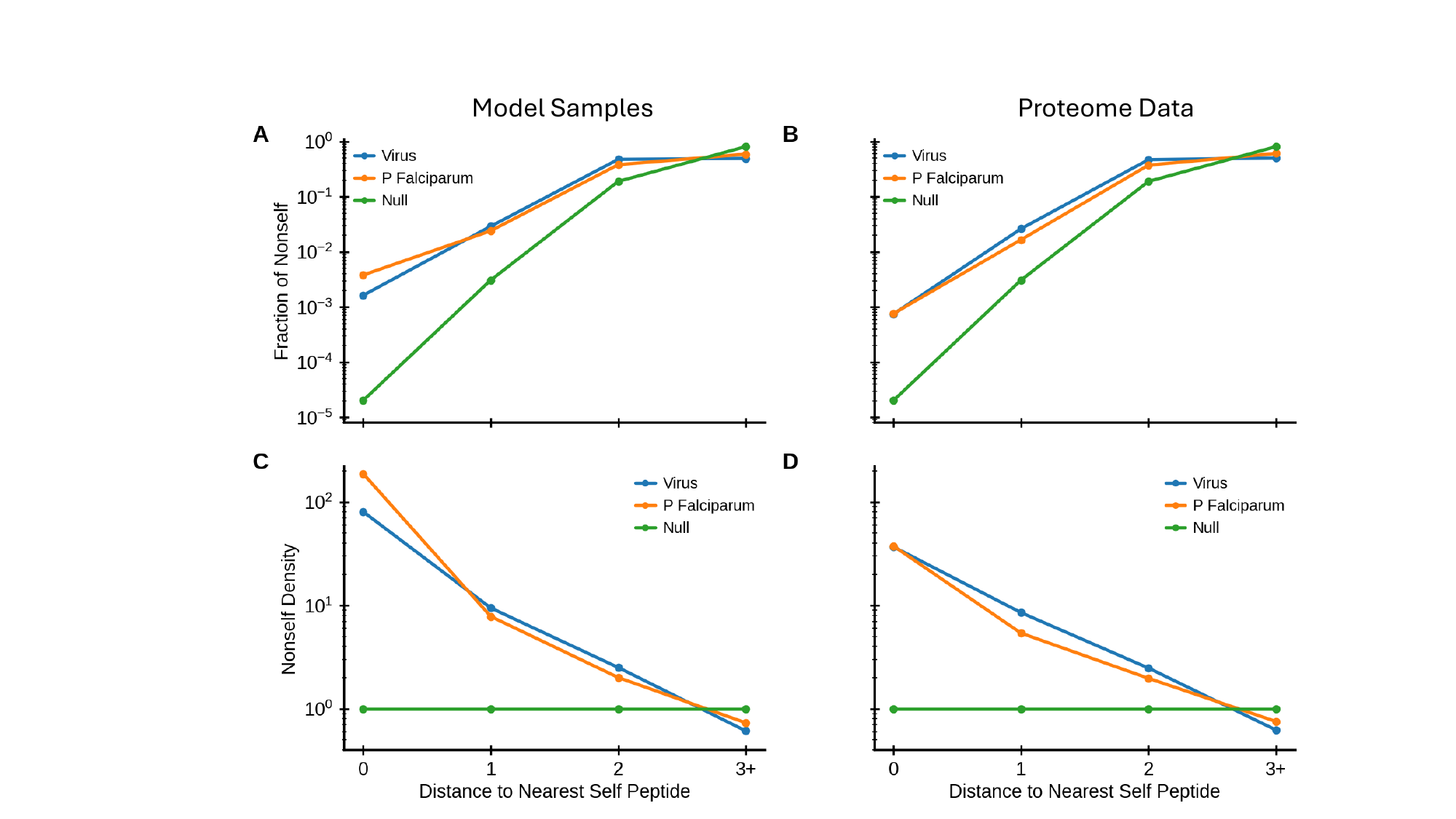}
	\end{center}
	\caption{
		\textbf{Distance to nearest self-peptide for model samples versus actual data}
		(A-B) Distribution of distances to the nearest self-peptide for peptides from samples drawn from models (A) and actual data (B). (C-D) Relative density of distances to the nearest self-peptide for the same sets of peptides. 
		Despite the P. Falciparum proteome having a relatively large KL divergence to the human proteome, its nearest neighbor distribution looks very similar to that of the human virome
		}
\end{figure*}
	
	\begin{figure*}[t!]
	\begin{center}
		\includegraphics[width=\textwidth]{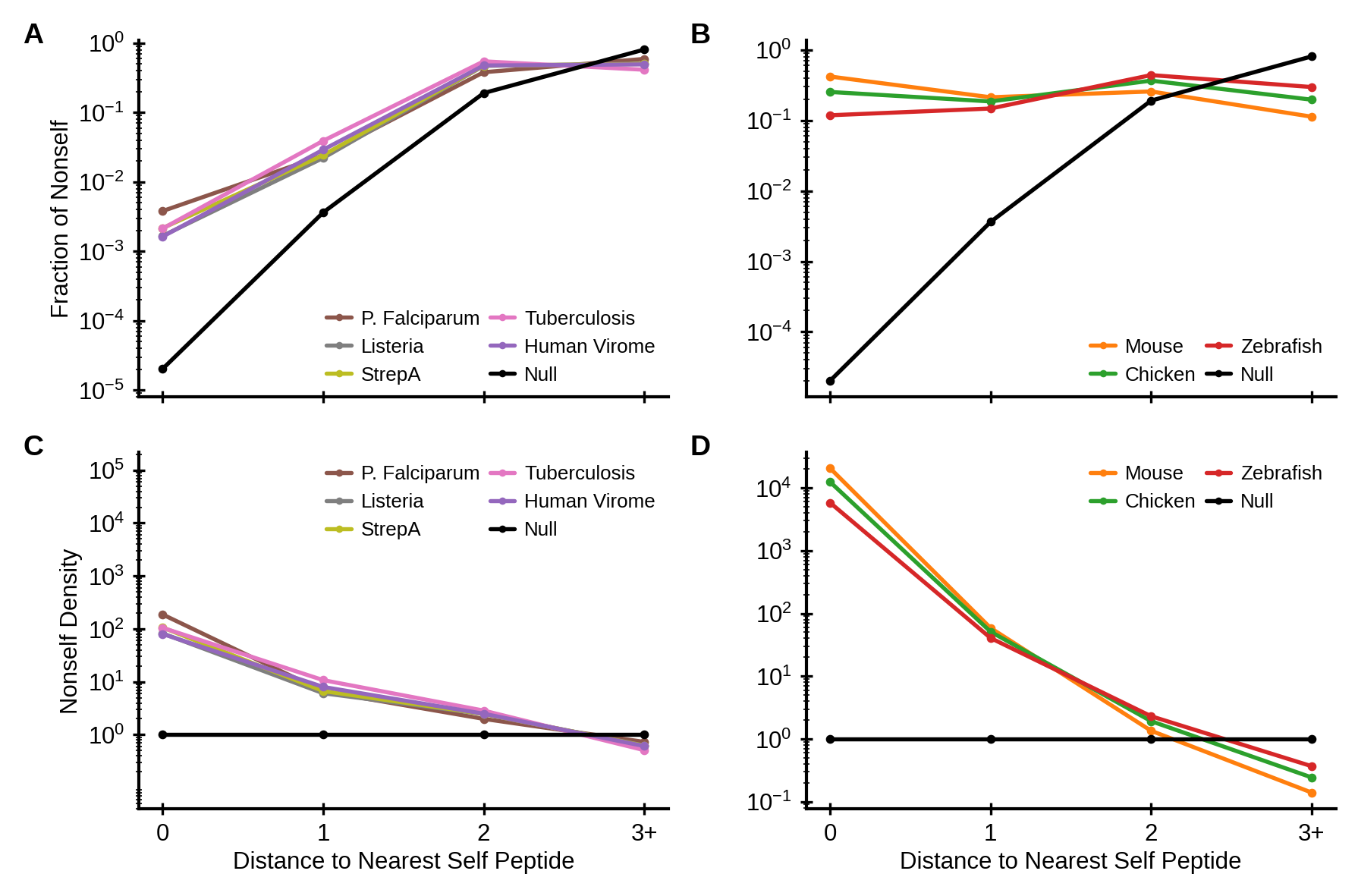}
	\end{center}
	\caption{
		\textbf{Distance to nearest self-peptide for different proteomes} 
		(A-B) Distribution of distances to the nearest self-peptide for peptides from different pathogen (A) and vertebrate (B) proteomes, and a null model with a uniform distribution over all 209 9–mers (black). (C-D) Relative density of distances to the nearest self-peptide for the same sets of peptides. Colors as in (A-B).
		The maximum entropy models are sufficient to closely predict the overlap between human and pathogen proteomes. However, the overlap between human and other vertebrate proteomes (C-D) is much higher due to the existence of many phylogenetically related proteins, which is not captured in our model. }
\end{figure*}
	\begin{figure*}[t!]
	\begin{center}
		\includegraphics[width=\textwidth]{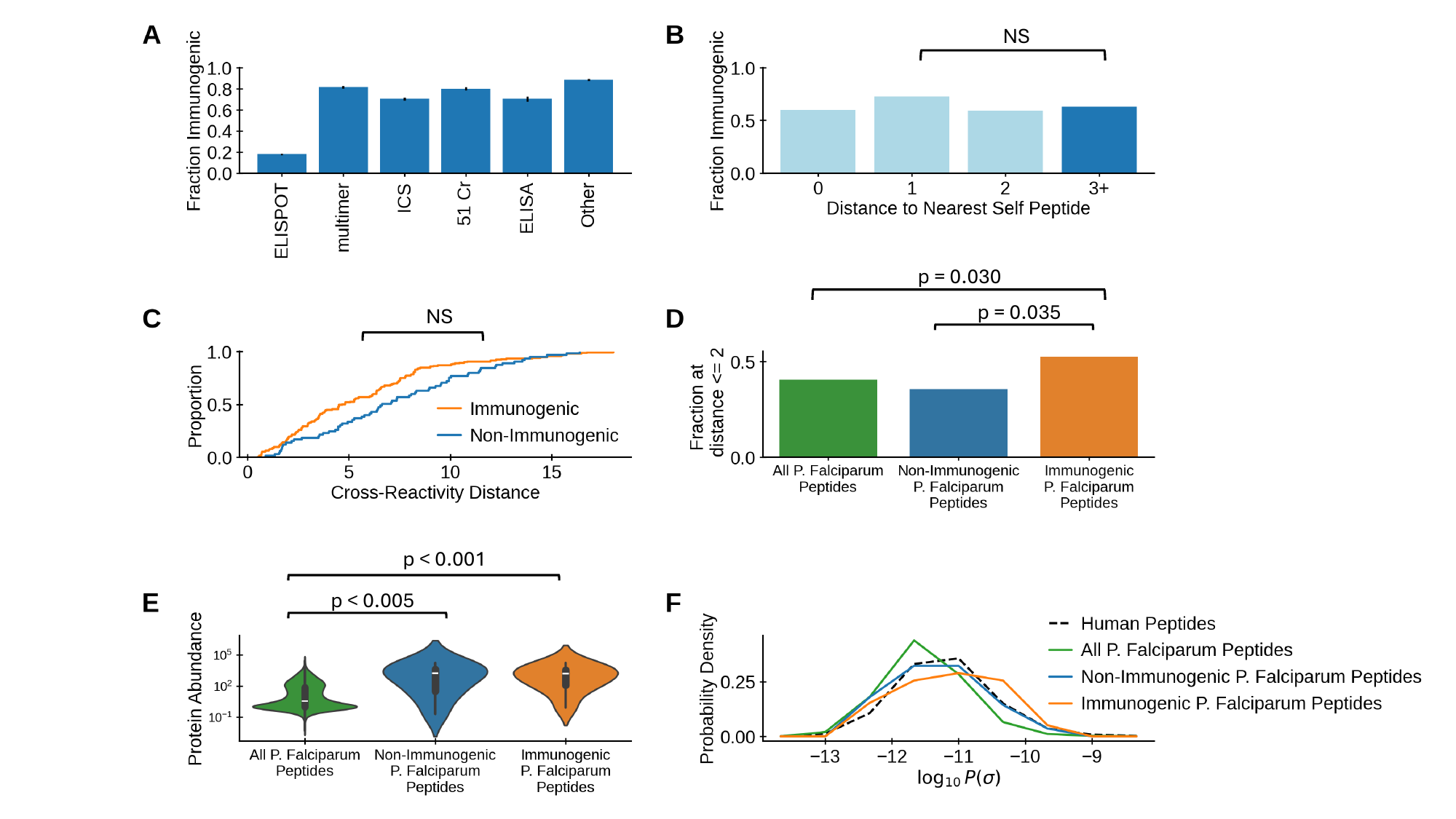}
	\end{center}
	\caption{
		\textbf{Immunogenicity for ELISPOT vs other assays} 
		(A) Fraction of peptides annotated to be immunogenic differs strongly between assays, potentially confounding analyses of immunogenicity. Assays: ELISPOT - enzyme-linked immune absorbent spot assay, multimer - pMHC multimer binding assay, ICS - intracellular cytokine staining assay, 51 Cr - Chromium-51 release assay, ELISA - enzyme-linked immunosorbent assay. We thus split the analysis into peptides measured via ELISPOT alone (Figure 4) and all available assays (B-F).
		(B) Immunogenicity (as ascertained using all available assays) as a function of Hamming distance to the nearest self-peptide. Immunogenicity did not differ significantly between distance bins ($\chi^2$ test).
		(C) Empirical cumulative distribution functions of T cell cross-reactivity distances\cite{Luksza2022} for immunogenic (orange) and non-immunogenic (blue) epitopes differing from the nearest self-peptide by a single amino acid. For hamming-distance-one peptides, cross-reactivity distances for immunogenic epitopes (as ascertained using all available assays) do not differ significantly from non-immunogenic epitopes (Kolmogorov–Smirnov Test).
		(D)  Fraction of different sets of peptides from the F. Falciparum proteome within distance 2 of the nearest self-peptide. Fraction with distance <= 2 differs between immunogenic peptides (orange), non-immunogenic peptides (blue), and all peptides (green) ($\chi^2$ test).
		(E) Average protein abundance of the source protein of different sets of peptides from the F. Falciparum proteome. Abundance significantly differed between the source proteins for IEDB epitopes (blue and orange) and the background distribution of all peptides (green). Abundance did not differ between immunogenic (orange) and non-immunogenic peptides (blue) (All assays, Mann-Whitney U-test).
		(F) Distribution of (log) probabilities given by the human proteome maximum entropy model for different sets of peptides from the F. Falciparum proteome.
		Data: Foreign epitopes tested by any assay as described in Methods. Protein abundance data from Pax-DB\cite{huang2023paxdb}(see Appendix D).}
\end{figure*}
	
		\begin{figure*}[t!]
		\begin{center}
			\includegraphics[width=\textwidth]{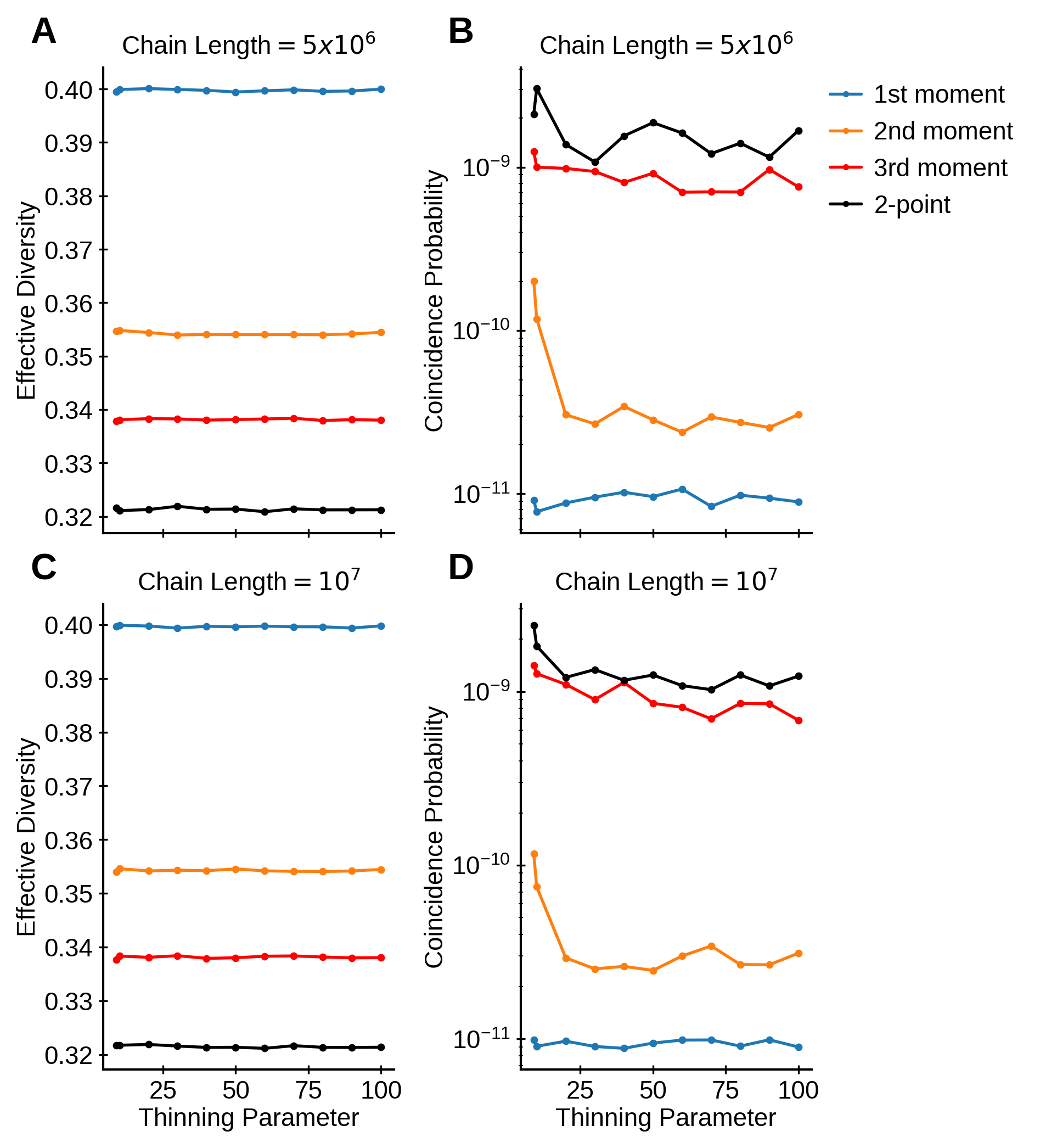}
		\end{center}
		\caption{
			\textbf{Statistics on samples are robust to hyperparameters} 
				Sampling from the models is used to estimate entropy and coincidence. Estimates of effective diversity (A, C) and coincidence probability (B, D) for different hyperparameters used to draw samples from the models using the Metropolis-Hastings Algorithm54,55 (see Methods). Data is shown here for the family of models trained on the human proteome. }
	\end{figure*}

\end{document}